# Terahertz electrodynamics in a zero-field Wigner crystal

Su-Di Chen[1,2,3,4*], Ruishi Qi[1,2,3], Ha-Leem Kim[1,2,3], Qixin Feng[1,2], Ruichen Xia[1], Dishan Abeysinghe[1,2], Jingxu Xie[1,2,5], Takashi Taniguchi[6], Kenji Watanabe[7], Dung-Hai Lee[1,2], Feng Wang[1,2,3*]

[1]Department of Physics, University of California, Berkeley, CA 94720, USA.

[2]Materials Sciences Division, Lawrence Berkeley National Laboratory, Berkeley, CA 94720, USA.

[3]Kavli Energy NanoScience Institute, University of California, Berkeley and Lawrence Berkeley National Laboratory, Berkeley, CA 94720, USA.

[4]Department of Physics, University of Florida, Gainesville, FL 32611, USA.

[5]Graduate Group in Applied Science and Technology, University of California, Berkeley, CA 94720, USA.

[6]Research Center for Materials Nanoarchitectonics, National Institute for Materials Science, 1-1 Namiki, Tsukuba 305-0044, Japan.

[7]Research Center for Electronic and Optical Materials, National Institute for Materials Science, 1-1 Namiki, Tsukuba 305-0044, Japan.

[*]Corresponding authors. Email: sudichen@ufl.edu, fengwang76@berkeley.edu

**In clean two-dimensional (2D) systems, electrons are expected to self-organize into a regular lattice—a Wigner crystal[1]—when their mutual Coulomb repulsion overwhelms kinetic energy. Understanding the Wigner crystal at zero magnetic field is a long-sought goal in physics, thanks to its fundamental simplicity and possible connection to the density-driven metal-insulator transition[2–6]. To date, evidence for such a crystal has been reported across various platforms[7–16]. However, the AC conductivity of a zero-field Wigner crystal, a key observable characterizing its electrodynamics, has never been measured. Here, we develop an ultrasensitive on-chip terahertz (THz) spectroscopy technique to probe the AC conductivity in electrostatically gated monolayer $MoSe_2$ encapsulated in hexagonal boron nitride. We observe a sub-THz resonance corresponding to the pinning mode of a zero-field Wigner crystal, whose frequency is orders of magnitude higher than those under high magnetic fields[17–19]. Using the pinning mode as an indicator, we reveal that moderate disorder notably stabilizes the Wigner crystal. With increasing density towards melting, we find that the pinning mode of the Wigner crystal coexists with a growing Drude component characteristic of an electron liquid, and the competition between these two components in the conductivity spectra leads to the insulator-metal transition of the 2D electron system. Our findings not only elucidate the low-energy electrodynamics of a zero-field Wigner crystal, but also establish on-chip THz spectroscopy as a powerful probe for correlated quantum phases in two-dimensional materials.**

As the carrier density (*n*) decreases, the ground state of a pristine 2D electron system at zero magnetic field is expected to evolve from a Fermi liquid to a Wigner crystal (Fig. 1a), when the ratio of Coulomb to kinetic energy, $r_s$, exceeds approximately 30–40 according to quantum Monte Carlo calculations[20,21]. In the presence of disorder, the crystal's sliding motion is suppressed; the formation of such a pinned Wigner crystal would subsequently lead to a metal-insulator transition. Despite persistent interest, experimentally investigating and understanding this evolution has been challenging, due to both the stringent requirements for low carrier density, temperature, and disorder, and the limited methods available for probing the internal structure of electron systems under such conditions. Semiconducting van der Waals (vdW) heterostructures, such as monolayer $MoSe_2$ encapsulated in hexagonal boron nitride (hBN), offer new opportunities to address these challenges[10,11,13–15]. Compared to conventional semiconductor platforms[8,9,12], the larger electron effective mass ($m^*$) in $MoSe_2$[22], combined with the lower dielectric constant of hBN[23], enables the system to reach the same $r_s$ at densities more than an order of magnitude higher ($n \sim 4\times10^{11}$ cm$^{-2}$ for $r_s \sim 30$). This allows the relevant physics to be explored over a much broader range of densities and temperatures, and renders the system more robust against moderate disorder. Indeed, optical evidence for Wigner crystallization has been reported in monolayer $MoSe_2$[10,13]. In the closely related bilayer $MoSe_2$, a Wigner solid has also been imaged using scanning tunneling microscopy (STM)[14]. Separately, DC transport measurements on the monolayer have observed a metal-insulator transition near $r_s \sim 30$[24], approaching theory predictions in the clean limit[20,21]. However, the AC conductivity of such systems, which can reveal the pinning mode resonance[25–27], a smoking-gun signature of Wigner crystallization, remains unexplored.

In magnetic-field-induced Wigner crystals[28–30], the pinning mode has been observed around GHz frequencies[17–19], arising from the interplay between the crystalline structure, disorder potential, and Lorentz force[25,26]. In zero-field Wigner crystals, the counterpart of this mode has long been predicted[27] but has yet to be observed. Traditionally, the pinning mode in field-induced Wigner crystals is detected by measuring microwave absorption through a meandered, centimeter-long transmission line capacitively coupled to the 2D electron system[17–19]. However, applying this technique to samples like $MoSe_2$ presents two major difficulties. First, the small lateral size of vdW heterostructures (~ 20 μm) limits the interaction length of the transmission line to about 1000 times smaller than that of conventional semiconductor devices, dramatically reducing absorption. Second, at zero field, the absence of Lorentz force will restore the pinning mode to its intrinsic frequency scale much higher than a few GHz, beyond the reach of standard microwave spectroscopy. Therefore, a measurement that is both ultrasensitive and broadband is needed.

**Ultrasensitive on-chip THz spectroscopy**

Here, we employ our newly developed technique[31] that directly measures the complex conductivity ($\sigma$) of electrostatically gated vdW semiconductors in the THz frequency range. Briefly, we capacitively couple the sample to the center section of a coplanar stripline (CPS) waveguide (Fig. 1b and c), inject THz pulses into the CPS using a photoconductive switch on one side, and detect the transmitted electric field in the time domain using a second switch on the opposite side (Extended Data Fig. 1). From the Fourier transform of time-domain traces at various gate voltages (Extended Data Fig. 2), we obtain the frequency-dependent complex transmission coefficient ($t$) normalized by the reference at zero density ($t_0$). The ratio $t/t_0$ is then converted to $\sigma$ of the sample, using numerical mappings obtained through electromagnetic simulations based on accurate device parameters (Extended Data Fig. 3).

Our recent proof-of-principle experiment[31] already achieves drastically higher sensitivity than earlier generations of on-chip THz measurements[32,33], detecting $\sigma$ down to 10 μS at 20 K. This is enabled by several key improvements. First, we utilize a tapered CPS design that significantly increases sample absorption while minimizing multireflection effects (Extended Data Fig. 1a). Second, by simultaneously monitoring the injected THz amplitude and detector responsivity in addition to the transmitted signal, we can normalize out extrinsic drifts and more accurately obtain the intrinsic transmission (Extended Data Fig. 1b). Third, the use of a thin indium tin oxide (ITO) gate electrode with few-fs transport scattering time allows us to separate the THz response of the sample from that of the gate electrode, ensuring background-free measurements (Extended Data Figs. 3 and 4). In this study, we have further optimized the fabrication of photoconductive switches to reduce noise, and rebuilt our experiment in an improved environment that enables continuous averaging over weeks at sample temperatures down to 4.5 K (Methods). With these improvements, we can now detect $\sigma$ down to 1 μS, reaching the regime relevant to Wigner crystal physics.

**THz resonance in lightly-doped $MoSe_2$**

We have studied two $MoSe_2$ devices with almost identical design but different levels of disorder, using monolayer $MoSe_2$ flakes exfoliated from two distinct batches of bulk crystals (Methods, Fig. 1b, and Extended Data Fig. 5). Fig. 1d presents the optical reflectance spectra from the first device (Device 1), which display sharp resonances from excitons and trions, or repulsive and attractive polarons. The exciton response allows us to determine the doping state of the sample (Methods and Extended Data Fig. 6), and to reproduce the previously reported optical evidence for Wigner crystal[10]: the periodic potential from the Wigner crystal lattice folds the exciton band structure, creating a weak Umklapp sideband at energy $E_x+h^2n/\sqrt{3}m_x$ in the reflectance spectra. Here, $E_x$ denotes the exciton energy, $h$ is the Planck constant, and $m_x$ is the exciton effective mass. The gate-voltage derivative of the reflectance spectra shows this sideband feature (Fig. 1e and Extended Data Fig. 7), which gradually diminishes with increasing density, consistent with earlier studies[10,13].

On the same device, our AC conductivity measurement uncovers a more pronounced anomaly in the THz frequency range. In Fig. 1f, we show the conductivity spectra (filled circles) at $2.0\times10^{11}$ cm$^{-2}$ ($r_s \sim 43$), a density where the Umklapp feature is unambiguously present. The spectra highlight a broad resonance around 0.5 THz, as indicated by the peak in the real part of conductivity ($\sigma_1$) and the corresponding zero crossing in the imaginary part ($\sigma_2$). The overall width of this resonance is consistent with the predicted pinning mode of a zero-field Wigner crystal (solid curves)[25–27], although differences exist in the detailed lineshape, which could result from the finite temperature in our measurement.

In contrast, at $1.6\times10^{12}$ cm$^{-2}$ ($r_s \sim 15$), a higher density where the Umklapp feature has long faded, we observe Drude-like conductivity spectra without the sub-THz resonance (Fig. 1g). The data as a function of frequency ($f$) can be well fit to a modified Drude formula: $\sigma = \frac{D}{2\pi^2}\frac{1}{\Gamma_D - if} - i\alpha f$, where $D$ is the Drude weight, $\Gamma_D$ is the scattering rate, and $\alpha$ is a non-negative constant approximately accounting for the background in $\sigma_2$ arising from higher-frequency absorptions outside the Drude peak. Consistent with the non-negligible $\alpha$, the extracted Drude weight is only 60% of that estimated using $m^* = 0.8\, m_0$, where $m_0$ is the free electron mass[22]. This suppression weakens at higher temperature and density, or with lower disorder (Extended Data Fig. 8), reflecting the role of disorder in promoting localization.

**Density and disorder dependence**

To understand the distinct features in the conductivity spectra, we show the detailed density evolution of $\sigma_1$ from Device 1 in Fig. 2a. Upon doping, $\sigma_1$ is initially dominated by the sub-THz resonance, whose frequency and amplitude both increase with density. Around $5\times10^{11}$ cm$^{-2}$, the resonance frequency ($f_p$) continues to rise, but its amplitude starts to decline. Meanwhile, in the low-frequency limit, a distinct Drude-like component gradually gains spectral weight, and is clearly observed to coexist with the resonance over an intermediate density range up to approximately $8\times10^{11}$ cm$^{-2}$. At higher densities, the resonance eventually disappears, and the spectra feature a single Drude peak, similar to that in Fig. 1g.

We observe qualitatively similar behaviors in the second device (Device 2, Fig. 2b). This device exhibits lower disorder, as evidenced by the much narrower Drude peak and an over sixfold increase in mobility at ~ $1\times10^{12}$ cm$^{-2}$ (Fig. 2b inset) compared to Device 1. Consequently, the full evolution seen in Fig. 2a now occurs over a significantly lower density range: the resonance weakens around $2\times10^{11}$ cm$^{-2}$ and vanishes near $4\times10^{11}$ cm$^{-2}$. Moreover, $f_p$ is also substantially lower than that in Device 1 at the same density.

While the Drude peak represents a metallic, Fermi-liquid-like state, the origin of the resonance warrants discussion. We first rule out plasmonic resonances set by the sample or CPS size[33,34], as they will not affect the measured conductivity spectra in the conductivity ranges studied here (see Methods and Extended Data Fig. 9 for more discussions), nor would they be restricted to lower densities in cleaner samples as our data show. Another possibility is plasmonic resonances from spontaneously formed metallic puddles. However, with increasing density these puddles would expand and eventually merge, causing the resonance to redshift and morph into the Drude peak, which is inconsistent with our observations.

Therefore, the resonance must originate from an insulating state. In an insulator driven by Anderson localization, $\sigma_1$ is expected to increase smoothly with frequency up to a soft gap edge, whose frequency decreases as the system approaches the insulator-metal transition with increasing density[35,36]. This behavior is again inconsistent with our observations. The remaining possibility is a pinned Wigner crystal, where resonances can arise from both single-particle excitations and the collective pinning mode. The frequencies of single-particle excitations are largely insensitive to disorder and predicted to lie well above our measurement window[37]. For the pinning mode, $f_p$ is expected to increase with disorder at fixed density[27], consistent with our experiment. We thus attribute the observed resonance to the pinning mode.

As such, the frequency-dependent conductivity measured here supports the Umklapp interpretation of the exciton sideband[10] and confirms the existence of a zero-field Wigner crystal. Moreover, the presence of both the pinning mode and the Drude component in the conductivity spectra at intermediate densities indicates the coexistence of electron solid and liquid regions near the melting of the Wigner crystal. This is consistent with recent optical[13] and STM[14] measurements.

Remarkably, the $f_p$ observed here in $MoSe_2$ far exceeds those in field-induced Wigner crystals[17–19]. While stronger disorder in $MoSe_2$ may contribute, the dominant factor is the Lorentz force, which under high magnetic fields alters the electron motion and reduces the pinning mode frequency to $f_{p,0}^2/f_c$, where $f_{p,0}$ is the pinning frequency without Lorentz force, and $f_c$ is the cyclotron frequency. As a typical example, for the high-field Wigner crystal in n-type GaAs, $f_{p,0}$ can reach around 100 GHz while $f_c \sim 6300$ GHz at 15 T[17]. A further distinction in $MoSe_2$ is that $f_p$ increases with density, opposite to the trend observed in Wigner crystals stabilized by high magnetic fields[18]. This may arise from the density-dependent electron wavefunction size (i.e., the electron's spatial spread at each site of the Wigner crystal) in zero-field Wigner crystals, which can enhance the effective disorder the electron sees at higher densities[27], distinct from Wigner crystals in a Landau level where the wavefunction size is fixed by the magnetic length.

**Quantitative aspects of density evolution**

We now examine the density evolution more quantitatively. In Fig. 3a and b, we plot the relative frequency slope of $\sigma_1$ well below $f_p$ for both devices. The slope is positive at low densities and becomes negative as the density increases. Since we expect positive (negative) slopes for insulating (metallic) states, the observed zero crossing provides an estimate for the location of the insulator-metal transition. The crossings are found at $r_s \sim 25(1)$ and $41(2)$ for devices 1 and 2, respectively. These values are generally comparable to theory predictions for the quantum melting of Wigner crystals[20,21]. Meanwhile, the difference between our two devices demonstrates that moderate disorder can further stabilize the Wigner crystal. This occurs because the solid phase, characterized by a spatially nonuniform charge density, can better adapt to local disorder to lower its energy, echoing recent theoretical proposals[6].

Next, we extract $f_p$ and the width ($\Gamma$) of the pinning mode by fitting $\sigma_1$ to a Lorentz oscillator model plus a Drude background (Extended Data Fig. 10): $\sigma = \frac{L}{2\pi^2}\frac{1}{\Gamma - i(f^2 - {f_p}^2)/f} + \frac{D}{2\pi^2}\frac{1}{\Gamma_D - if}$, where $L$ represents the spectral weight of the pinning mode. For both devices, $\Gamma/f_p$ remains close to unity (Fig. 3c inset) over the full density range within which the pinning mode is identifiable. Meanwhile, $f_p$ is found to scale with $n$ with a power exponent around 0.38 throughout most of this range (Fig. 3c). The data only deviate upward from the scaling behavior near the lowest densities, and we discuss the possible origins in detail in Methods.

Interestingly, the measured exponent in the power-law scaling is, within experimental uncertainties, consistent with the predicted value of 3/8 for zero-field Wigner crystals when the disorder correlation length is smaller than the electron wavefunction size[27]. The behavior of $\Gamma/f_p$ also agrees with theory[25–27]. However, we caution that the theoretical scaling from Ref.[27] is valid only deep within the crystalline phase. Furthermore, a quantitative estimation following the theory suggests that the parameters in our devices lie outside the collective pinning regime, within which the 3/8 scaling is derived. It is important to note that the theory assumes a pure Wigner crystal phase without considering the spatial inhomogeneity caused by the coexistence of Wigner crystal and electron liquid regions. In addition, it assumes a single disorder correlation length while in our devices, disorder associated with, e.g., atomic defects, trapped charges, and strain variations can exhibit multiple correlation lengths. Regardless of the precise mechanisms involved, our observations here provide an important benchmark for future studies.

**Temperature dependence**

Last, we study the temperature ($T$) evolution of conductivity. We focus on Device 1, where the higher energy scales make the measurements less challenging. Fig. 4a shows representative $\sigma_1$ spectra in the low-density regime. Here, the pinning mode, which is the dominant feature, shifts slightly towards lower frequencies and broadens as temperature increases. This causes a rise of $\sigma_1$ with $T$ at frequencies well below $f_p$, i.e., an insulating behavior. In contrast, at $1\times10^{12}$ cm$^{-2}$ (Fig. 4b), the low-temperature $\sigma_1$ spectrum only consists of the Drude peak, which also broadens upon warming, leading to the decrease of $\sigma_1$ in the low-frequency limit and thus metallicity.

At intermediate densities, the temperature evolution is found to be the combination of the two (Fig. 4c): the pinning mode and Drude peak coexist at low temperatures; with increasing density, when the Drude peak overtakes the pinning mode as the primary component, the low-frequency behavior changes from insulating to metallic. From the temperature dependence of $\sigma_1$ at 0.2 THz (Fig. 4d), we observe that the insulator-metal transition occurs around $r_s \sim 26$, consistent with our earlier estimation using the frequency slope of $\sigma_1$ at 4.5 K (Fig. 3a). Essentially, the frequency slope reflects whether the pinning mode or the Drude peak is the leading component, which in turn governs the temperature evolution. Therefore, our observed frequency, density, and temperature dependence of conductivity directly ties the disorder-pinned Wigner crystal to the insulating phase in the density-driven insulator-metal transition in monolayer $MoSe_2$.

In summary, we have measured the low-energy electrodynamics of a zero-field Wigner crystal for the first time, directly revealing its pinning mode and connection to the insulator-metal transition. The ultrasensitive THz technique developed here, capable of detecting μS-level conductivity in μm-sized samples, also opens up new possibilities for exploring other correlated quantum phases in vdW heterostructures.

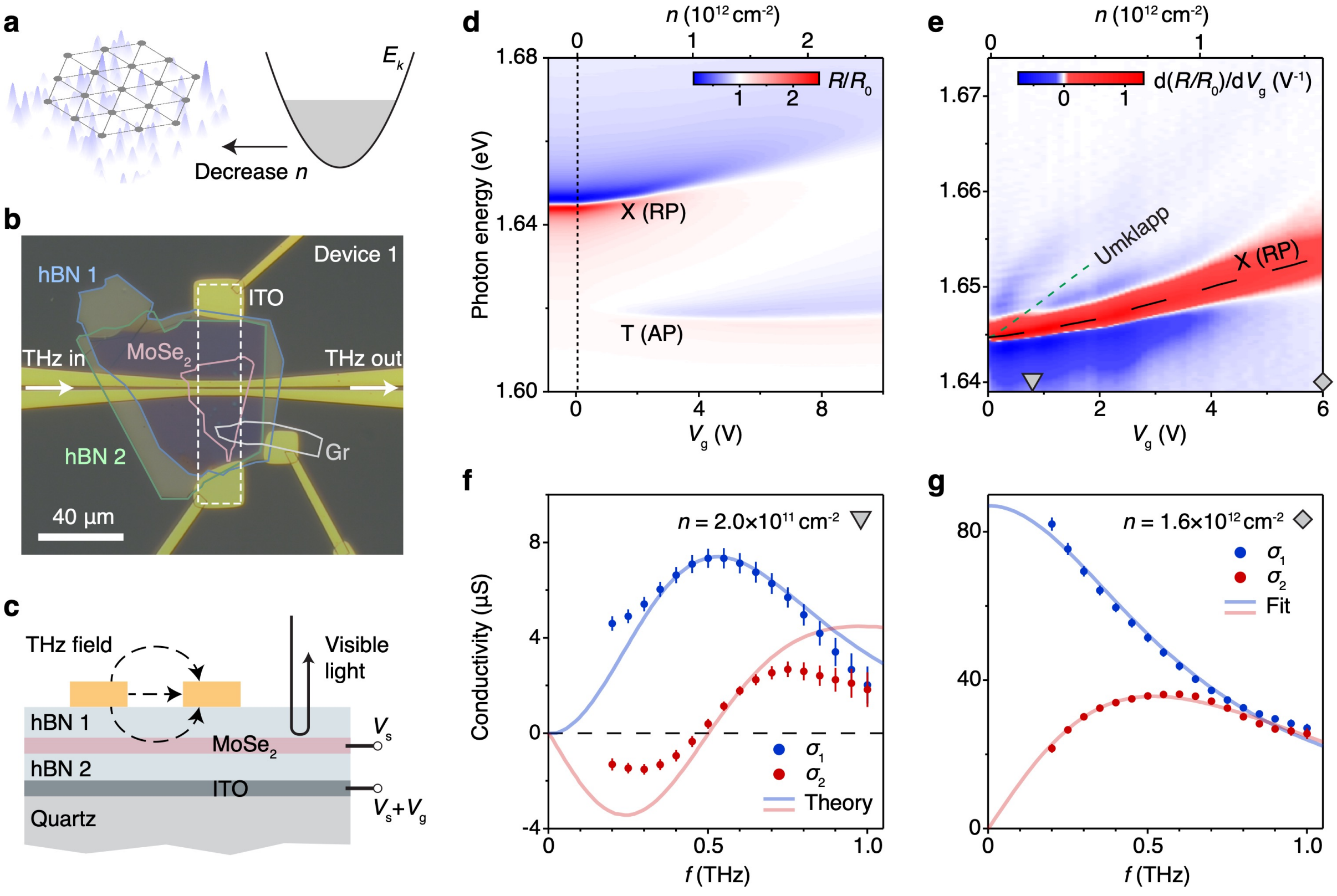

**Fig. 1 | Optical and THz spectroscopy on the same vdW stack. a**, Cartoon illustrations of a disorder-pinned Wigner crystal (left) and a Fermi liquid with band dispersion $E_k$ (right). **b**, Optical micrograph showing center section of the device. Monolayer $MoSe_2$, few-layer graphite contact (Gr), and hBN flakes are outlined in solid. White dashed box marks the indium-tin-oxide (ITO) backgate which defines an active region in $MoSe_2$. **c**, Cross-sectional schematic of the active region in **b** perpendicular to the THz propagation direction. Sample conductivity ($\sigma=\sigma_1+i\sigma_2$) is obtained by measuring the THz transmission coefficient through the waveguide (yellow) normalized to the zero-density reference. Optical reflectance ($R$) is measured on the same stack as illustrated by the solid arrow. **d**, Density ($n$) evolution of the normalized optical reflectance ($R/R_0$) spectra at 7 K. $R_0$ is measured off the $MoSe_2$ flake. X (RP) and T (AP) indicate the exciton (repulse polaron) and trion (attractive polaron) resonances, respectively. Vertical dashed line marks the onset of electron doping determined from the spectra (Extended Data Fig. 6). **e**, Derivative of data in **d** with respect to the gate voltage $V_g$, shown in smaller range. Black dashed curve shows the fitted energy of the exciton resonance ($E_x$); Green dashed curve represents the expected energy of the Umklapp sideband in a Wigner crystal, $E_x+h^2n/\sqrt{3}m_x$, with

$m_x = 1.4\ m_0$ (Extended Data Fig. 7). This sideband manifests as the white patch located between blue strips in the data. **f**. Representative conductivity spectra (circles) in the low-density regime, measured at $V_g = 0.8$ V (triangle in **e**) and 4.5 K. Real and imaginary parts are plotted in blue and red, respectively. Semi-transparent curves show the theoretical pinning-mode lineshape in a zero-field Wigner crystal at zero temperature from Ref.[26], scaled proportionally to match the peak position and height in the measured $\sigma_1$. **g**. Conductivity spectra at $V_g = 6$ V (diamond in **e**) and 4.5 K. Semi-transparent curves represent a simultaneous fit to both the real and imaginary parts using the modified Drude expression $\sigma = D/[2\pi^2(\Gamma_D - \mathrm{i}f)] - \mathrm{i}\alpha f$ with non-negative $\alpha$. Error bars in **f** and **g** represent 1σ uncertainty from averaging multiple measurements.

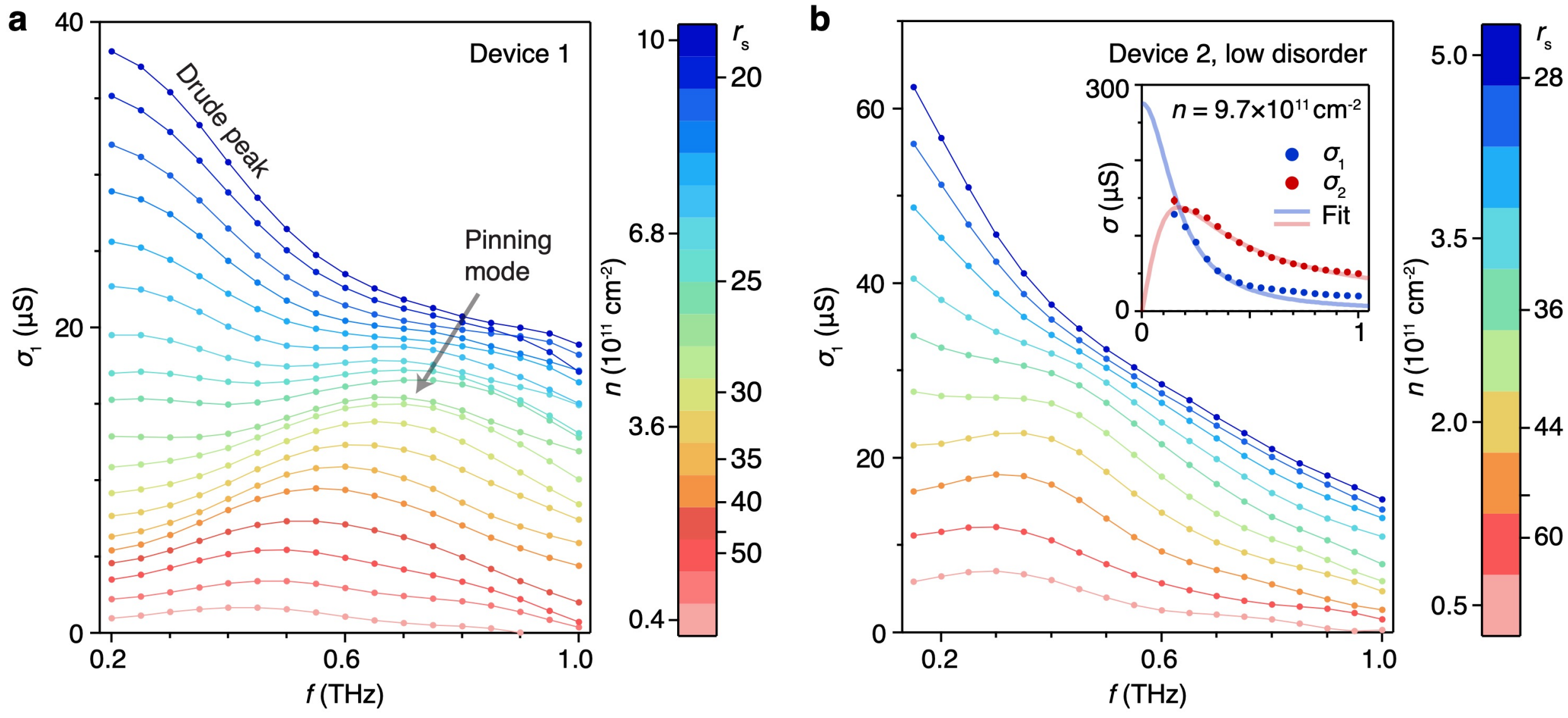

**Fig. 2 | Density and disorder dependence of AC conductivity. a**, Density evolution of conductivity spectra in Device 1 at 4.5 K. Only $\sigma_1$ is plotted for clarity. Curves are measured from $V_g$ = 0.2 V to 3.8 V in 0.2 V steps, with their corresponding densities and $r_s$ indicated by the color bar. The arrow highlights the pinning mode, which shifts upward in frequency with doping, and coexists with a low-frequency Drude component at intermediate densities. **b**, Same as **a** but for Device 2 at 4.8 K, measured from $V_g$ = -0.1 V to 1.7 V in 0.2 V steps. Electron doping onsets at $V_g$ = -0.3 V for this device. Inset: complex conductivity spectra at $V_g$ = 3.6 V, where the density roughly matches that of the topmost curve in **a**. Error bars represent 1σ uncertainty from averaging multiple measurements. Semi-transparent curves are fits to the modified Drude expression used in Fig. 1g. The narrower Drude peak and higher low-frequency conductivity in the inset indicate weaker disorder compared with Device 1, which leads to lower pinning frequencies in the lightly doped regime.

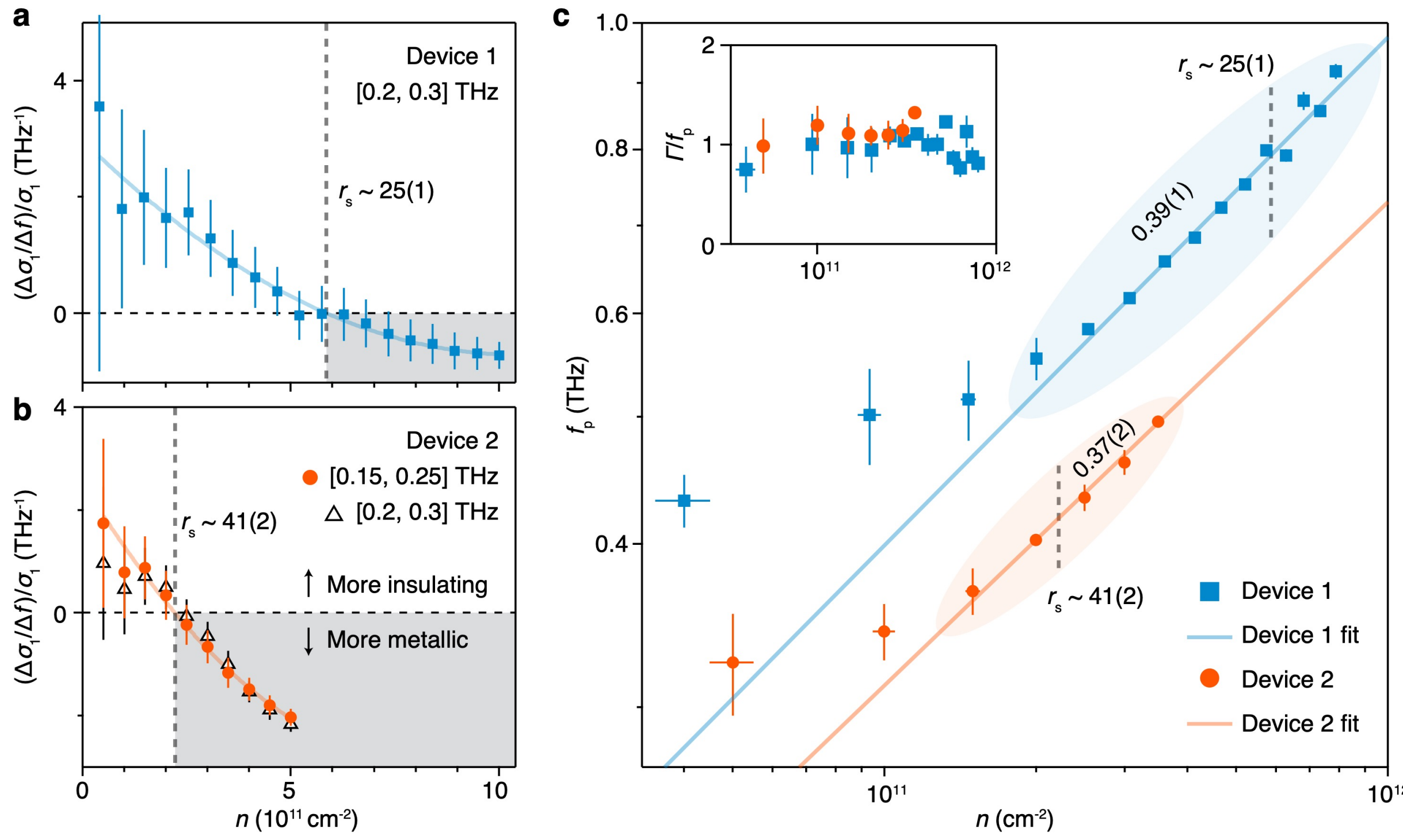

**Fig. 3 | Quantitative analysis of density evolution. a**, Relative slope of $\sigma_1$ in the low-frequency regime from Device 1 (squares). The slope is estimated using data from Fig. 2a, as $[\sigma_1(f_2)-\sigma_1(f_1)]/(f_2-f_1)$, where $f_1$ = 0.2 THz and $f_2$ = 0.3 THz, and normalized by $[\sigma_1(f_2)+\sigma_1(f_1)]/2$. A quadratic fit (curve) crosses zero at $5.9(5)\times10^{11}$ cm$^{-2}$ (vertical dashed line), approximately marking the density of insulator-metal transition. **b**, Same as **a** for Device 2. The relative slope is calculated between 0.15 and 0.25 THz (circles) using data from Fig. 2b and the zero crossing is found at $2.2(2)\times10^{11}$ cm$^{-2}$. The slope between 0.2 and 0.3 THz is also plotted (triangles) for completeness and shows consistent behavior. Error bars in **a** and **b** are propagated from 1σ uncertainties from averaging multiple conductivity measurements. **c**, Log-log plot of pinning frequency $f_p$ versus $n$ (markers). $f_p$ is extracted by fitting $\sigma_1$ in Fig. 2 to a Lorentz oscillator representing the pinning mode plus a Drude background (Extended Data Fig. 10). Shaded ellipses highlight power-law behavior observed in both devices except at very low densities. Semi-transparent lines are linear fits to the log-log data inside the ellipses, revealing a power exponent of 0.39(1) for Device 1 and 0.37(2) for Device 2. Vertical dashed lines mark the zero-crossing densities from **a** and **b**. Inset: linear-log plot of width ($\Gamma$)-to-$f_p$ ratio of the pinning mode versus $n$. Vertical error bars are propagated from 1σ fitting error. Horizontal error bars reflect uncertainties in determining the doping onset (Extended Data Fig. 6).

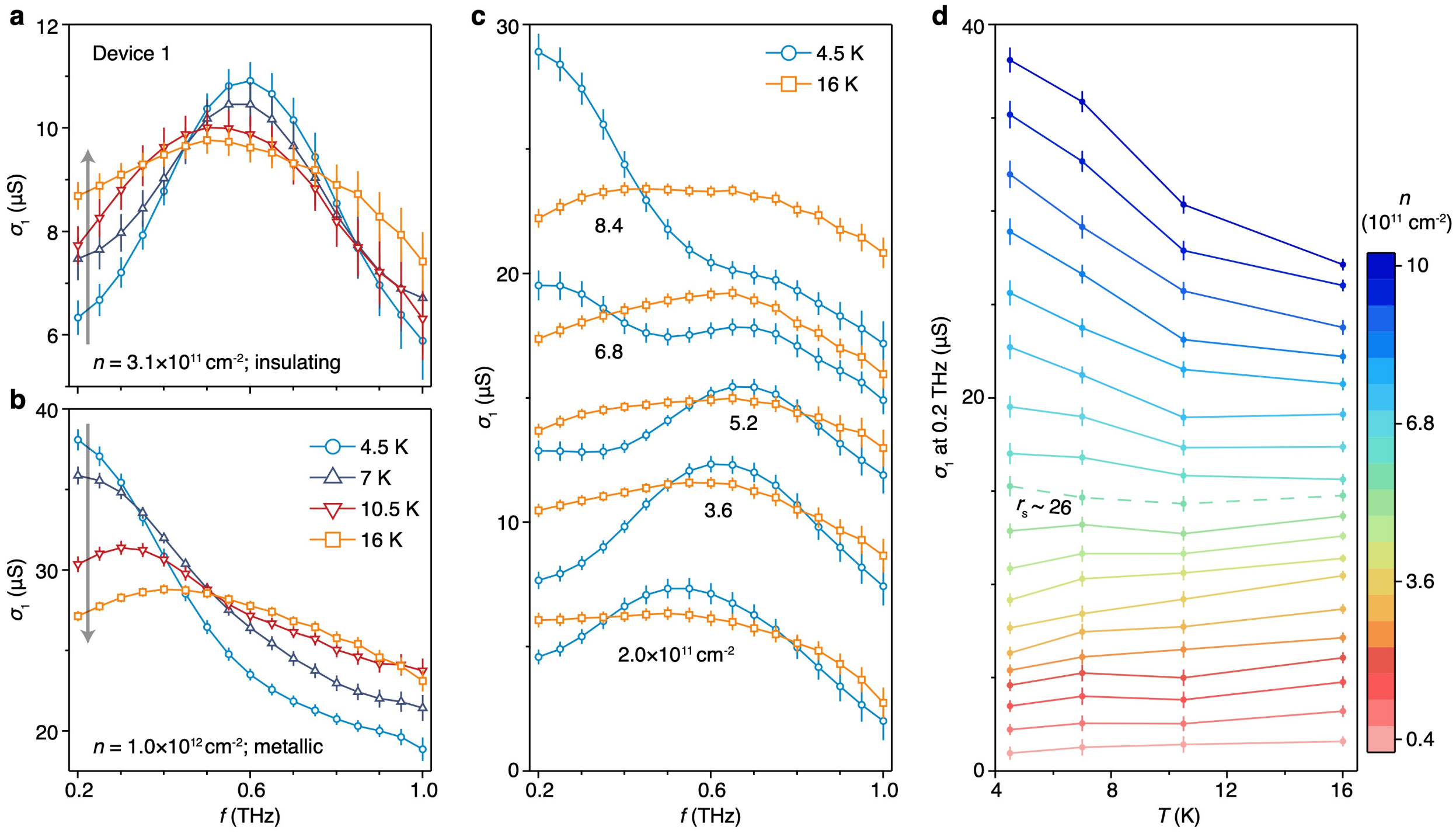

**Fig. 4 | Temperature dependence of AC conductivity. a**, Representative $\sigma_1$ spectra in the low-density regime, measured at $V_g = 1.2$ V in Device 1. Grey arrow highlights the low-frequency effect caused by the redshift and broadening of pinning mode from 4.5 K to 16 K. **b**, Same as **a** at $V_g = 3.8$ V in the metallic regime, where the Drude peak broadens with temperature. **c**, $\sigma_1$ spectra across the insulator-metal transition, from $V_g = 0.8$ V to 3.2V in 0.6 V steps. The corresponding density is labeled underneath each pair of spectra. The low-frequency behavior changes from insulating to metallic as the Drude peak overtakes the pinning mode at 4.5 K with increasing density. **d**, $\sigma_1$ at 0.2 THz, the lowest frequency measured in Device 1, plotted as a function of temperature ($T$). Curves are measured from $V_g = 0.2$ V to 3.8 V in 0.2 V steps, with their corresponding densities indicated by the color bar. The dashed curve highlights the onset of metallic behavior at $5.7\times10^{11}$ cm$^{-2}$, where the conductivity increases with decreasing temperature. All error bars represent 1σ uncertainty from averaging multiple measurements.

**Acknowledgements**

We thank K.-S. Kim, T. Wang, S.-K. Mo, Z.-X. Shen, and Y. He for discussions and Q. Wang for help with device fabrication. The measurements were supported by the NSF award no. 2311205. The device fabrication was supported by the U.S. Department of Energy, Office of Science, Office of Basic Energy Sciences, Materials Sciences and Engineering Division under contract no. DE-AC02-05-CH11231 (van der Waals heterostructures program, KCWF16). The simulations were performed at the Molecular Graphics and Computation Facility (MGCF) at UC Berkeley and the MGCF is in part supported by NIH S10OD034382. K.W. and T.T. acknowledge support from the JSPS KAKENHI (Grant Numbers 21H05233 and 23H02052), the CREST (JPMJCR24A5), JST and World Premier International Research Center Initiative (WPI), MEXT, Japan. D.-H.L. is supported by the U.S. Department of Energy, Office of Science, Office of Basic Energy Sciences, Materials Sciences and Engineering Division under contract no. DE-AC02-05-CH11231 within the Quantum Materials Program (KC2202). R.Q. and H.-L.K. acknowledge support from the Kavli ENSI Graduate Student Fellowship. S.-D.C. acknowledges support from the Kavli ENSI Heising-Simons Junior Fellowship.

**Author contributions**

F.W. and S.-D.C. proposed the experiment; S.-D.C., H.-L.K, Q.F., and R.X. fabricated the devices with help from J.X.; S.-D.C. performed the THz measurement with help from R.Q. and D.A.; S.-D.C. and R.Q. performed the optical measurement; S.-D.C. analyzed the data with input from D.-H.L. and F.W.; T.T. and K.W. grew the hBN crystals; S.-D.C. and F.W. wrote the manuscript with input from all authors; F.W. supervised the project.

**Materials & Correspondence**

Correspondence and requests for materials should be addressed to S.-D.C. or F.W.

**Competing interests**

The authors declare no competing interests.

**Data availability**

The data that support the findings of this study are available from the corresponding authors upon reasonable request.

## Methods

### Device fabrication

Our device fabrication largely follows the procedure detailed in Ref.[31] with a few modifications. $MoSe_2$ (purchased from HQ graphene), hBN, and graphite flakes are first exfoliated onto $SiO_2$/Si substrates. Using a bisphenol-A polycarbonate (PC) stamp, we sequentially pick up the top hBN, few-layer graphite (Gr), monolayer $MoSe_2$, and bottom hBN flakes, and release the assembled stack onto a prepatterned ITO gate electrode on a quartz substrate. A first photolithography step, followed by Cr/Au deposition, defines contacts to Gr and ITO, as well as the wide sections of the CPS waveguide. We measure the actual CPS pattern using an optical microscope to obtain geometrical parameters for subsequent steps.

Low-temperature-grown GaAs flakes, prepared using methods from Ref.[32], are then transferred to designated positions in the CPS. A second photolithography and Cr/Au deposition step connects the GaAs flakes to the CPS as the emitter and detector. Here, the photoresist is overdeveloped deliberately and descumed in oxygen plasma to minimize surface residue. Moreover, immediately before metallization, the chip is immersed in 20% HCl to remove the native oxide layer on GaAs. These procedures improve the signal-to-noise ratio in low-temperature measurements.

Finally, a single round of electron-beam lithography and Cr/Au metallization defines the tapered CPS section on both the actual device and a reference chip, using identical process parameters. The precise dimensions of the tapered CPS are measured on the reference chip using scanning electron microscopy (SEM) and atomic force microscopy (AFM). The hBN thicknesses on the actual device are also measured using AFM. These experimentally measured geometries, rather than the nominal design values, are then used in simulations (see Ref.[31] for details) to establish the mapping between $t/t_0$ and $\sigma$.

For Device 1, the CPS, top hBN, and bottom hBN thicknesses are 184 nm, 65 nm, and 58 nm, respectively. The ITO length along the CPS direction is 20 μm. The width of each CPS trace is 22.7 μm and tapered down to 3.14 μm near the sample. The CPS gap is 5.5 μm and tapered down to 0.78 μm, which is still much larger than the relevant intrinsic length scales in $MoSe_2$. For Device 2, the corresponding values are 187 nm (CPS thickness), 71 nm (top hBN), 62 nm (bottom hBN), 26 μm (ITO length), 22.6 μm (CPS trace), 2.55 μm (CPS trace tapered), 5.3 μm (CPS gap), and 0.84 μm (CPS gap tapered).

From Fig. 1g, the DC mobility of Device 1 is estimated to be $3.4\times10^2$ cm$^2$/(V·s) at $1.6\times10^{12}$ cm$^{-2}$ and 4.5 K. For Device 2, we estimate a mobility around $1.8\times10^3$ cm$^2$/(V·s) at $9.7\times10^{11}$ cm$^{-2}$ and 4.8 K (Fig. 2b inset). While the mobility is expected to increase further with density, in our current devices the THz measurements are less reliable in the high-mobility regime (see Methods subsection “Evaluating the effects of finite-size plasmonic resonances”). The variation between devices likely originates from the batch-to-batch differences in the $MoSe_2$ bulk crystals.

**Measurements**

All measurements are performed in an optical cryostat (Quantum Design OptiCool) in a stable lab environment with 0.1°C temperature variation. For THz measurements, either a 5× Mitutoyo objective or a lens with 40-mm focal length (Thorlabs AC254-040-B-ML) is used to achieve a large field of view covering both photoconductive switches. The switches are excited at 80 MHz repetition rate with few-mW optical power from a tunable femtosecond Ti:sapphire laser (Coherent Chameleon). The wavelength is set to 790 nm to avoid possible absorption of scattered light at the sample position by $MoSe_2$. The measurement circuits are shown in Extended Data Fig. 1. Typical emitter photocurrent $I_{inj}$ ranges from 150 nA to 250 nA and we do not observe any nonlinear effects within this range.

The base temperature of our setup is 1.6 K as measured near the chip carrier. However, during THz measurements, the laser beams on the switches far away from the sample create a continuous-wave (CW) heating effect. To determine the actual sample temperature, we use ITO as an in-situ sensor after calibrating its resistance as a function of temperature without laser excitation (Extended Data Fig. 4). Decreasing the laser power would reduce heating but also degrade the signal-to-noise ratio. Thus, the lowest temperature achieved here (4.5 K) represents a balance between minimizing heating and maintaining sensitivity.

To facilitate the electrostatic gating of $MoSe_2$ at low temperature, the sample is illuminated using a light-emitting diode (Thorlabs M617L3) with an estimated optical power of 0.1 nW/μm$^2$. We have verified that changing this power by a factor of 2 does not affect our results.

For Device 1, the measured time-domain data containing non-zero signals are around 6 ps long (Extended Data Fig. 2a). This sets the lowest measurable frequency, as well as the frequency resolution, at approximately 0.17 THz. We thus exclude data below this value. For Device 2, slightly longer time-domain traces (7 ps with non-zero signals) are collected. The corresponding lowest measurable frequency and frequency resolution are around 0.14 THz. We therefore exclude data below 0.14 THz for this device. For both devices, frequencies above 1.0 THz are excluded due to the diminishing signal-to-noise ratio.

For optical measurements, a 20× Mitutoyo objective is used. A supercontinuum laser (Fianium Femtopower 1060) is focused to a ~ 2 μm spot on the sample. The incident power is lower than 50 nW. When necessary, the entire sample is also weakly illuminated with a halogen lamp to facilitate electrostatic gating. The reflected light from the laser spot is selected using a pinhole in confocal geometry and collected by a spectrometer with a thermoelectrically cooled camera (Princeton Instruments PIXIS 256E). The objective is moved to access the reference spot off the $MoSe_2$ flake. The fast oscillations in $R/R_0$ as a function of wavelength, due to the imperfect cancelation of interference fringes, are filtered out by removing Fourier components above 2.5 nm$^{-1}$. Each $R/R_0$ spectrum is further normalized such that its average value between 777 and 782 nm, which corresponds to an energy window far below the trion resonance, equals 1.

**Determining density and $r_s$**

The electron density in $MoSe_2$ is calculated as $n = \varepsilon_0\varepsilon_r(V_g\text{-}V_{g,onset})/(ed)$, where $\varepsilon_0$ is the vacuum permittivity, $\varepsilon_r = 2.8$ is the DC dielectric constant of hBN along the out-of-plane direction[31], $V_{g,onset}$ is the onset gate voltage for electron doping determined from exciton response[38] (Extended Data Fig. 6), $e$ is the elementary charge, and $d$ is the thickness of the bottom hBN dielectric. For both devices, $d$ is sufficiently large such that including quantum capacitance would change $n$ by less than 3%. This effect is thus ignored. While the proportionality between ($V_g$-$V_{g,onset}$) and $n$ holds well for each device, the absolute value of $n$ has an uncertainty around 10% dominated by the uncertainty in $\varepsilon_r$.

We calculate $r_s$ using $r_s = m^* e^2/(4\pi\varepsilon_0\varepsilon\hbar^2\sqrt{\pi n})$, where $m^* = 0.8\ m_0$ according to Shubnikov–de Haas oscillation measurements at high densities[22], $\varepsilon = 4.4$ is the effective dielectric constant estimated using the geometrical mean of the in-plane and out-of-plane dielectric constants of hBN[23], and $\hbar$ denotes the reduced Planck constant.

**Evaluating the effects of finite-size plasmonic resonances**

Here we provide a detailed explanation on why the conductivity spectra reported here are unaffected by plasmonic resonances set by the sample or CPS size.

In our work, we extract the sheet conductance $\sigma$ of the monolayer $MoSe_2$ from the experimentally measured $t/t_0$, using the simulated mapping between $\sigma$ and $t/t_0$. As long as the simulations incorporate the actual device geometry, the finite-size plasmonic effects in experiments are automatically included in the simulated mapping. As a result, the extracted $\sigma$ simply represents the intrinsic material property of $MoSe_2$ and in principle does not carry any information related to the sample or CPS geometry.

In practice, there is one approximation in our simulations to reduce computation cost: in the gated region, our model uses the average width of the $MoSe_2$ flake while the actual width varies across different cross-sections. Nevertheless, we have verified that in the $\sigma$ range relevant to this work, the simulated $t/t_0$ is not sensitive to changes in $MoSe_2$ width (Extended Data Fig. 9a and b), which validates our approximation.

This insensitivity can be understood as follows. Considering the cross-sectional view of our device (Fig. 1c), a signal at frequency $f$ inside the CPS can excite a plasmonic wave that propagates to the sample edge and reflects back. The detailed location of the edge will not affect our measurement if the wave is strongly damped during this process. For a single 2D sheet with conductance $\sigma$ at frequency $f$, the plasmon wavevector[39] $q = i4\pi\varepsilon_0\varepsilon_{eff}f/\sigma$, and the relative amplitude after propagating over a distance of $l$ is $\exp[-\mathrm{Im}(q)\cdot l] = \exp[-4\pi\varepsilon_0\varepsilon_{eff}fl\cdot \mathrm{Re}(\sigma)/|\sigma|^2]$. Here $\varepsilon_{eff} = (1+ \varepsilon_{sub})/2$ with $\varepsilon_{sub}$ denoting the substrate dielectric constant. Since the monolayer $MoSe_2$ is coupled to the ITO gate in our device, we can use the sum of their sheet conductance, $\sigma_1+\sigma_{ITO}+i\sigma_2$, to estimate the decay of cross-sectional plasmonic waves. As shown in Extended Data Fig. 9c and d, the relative amplitude after a mere 10 μm propagation is already negligible for the conductivity values studied in this work, consistent with the lack of sample-width dependence in $t/t_0$ in our simulations.

In summary, finite-size effects can arise from two sources: the CPS structure and the sample's outer edges. Effects associated with the CPS are inherently captured in our simulated mapping between $\sigma$ and $t/t_0$, while those associated with the outer edges are negligible across the frequency and $\sigma$ ranges studied here thanks to sufficient damping. We can therefore extract the intrinsic conductance of the sample unaffected by geometrical effects.

**Possible origin of the low-density deviation of $f_p$ from power-law scaling**

Here we discuss the possible cause of the deviation of $f_p$ from power-law scaling in the low-density limit (Fig. 3c). One possibility stems from the fact that in real devices there are multiple sources of disorder that pins the Wigner crystal, such as atomic defects, trapped charges, and strain variations, which can have different disorder correlation lengths $\xi_d$. In theory, the pinning physics is sensitive to the ratio between $\xi_d$ and the electron wavefunction size $\xi_e$. As $\xi_e$ lengthens with decreasing density, the disorder component most effective for pinning may change, causing the system to enter a different pinning regime. A second possibility is the formation of mesoscopic charge puddles at very low densities, where the actual electron density within the puddles exceeds the spatial average. The exciton response does not contradict this scenario: in this regime, the Umklapp sideband is too close to the exciton resonance to be resolved.

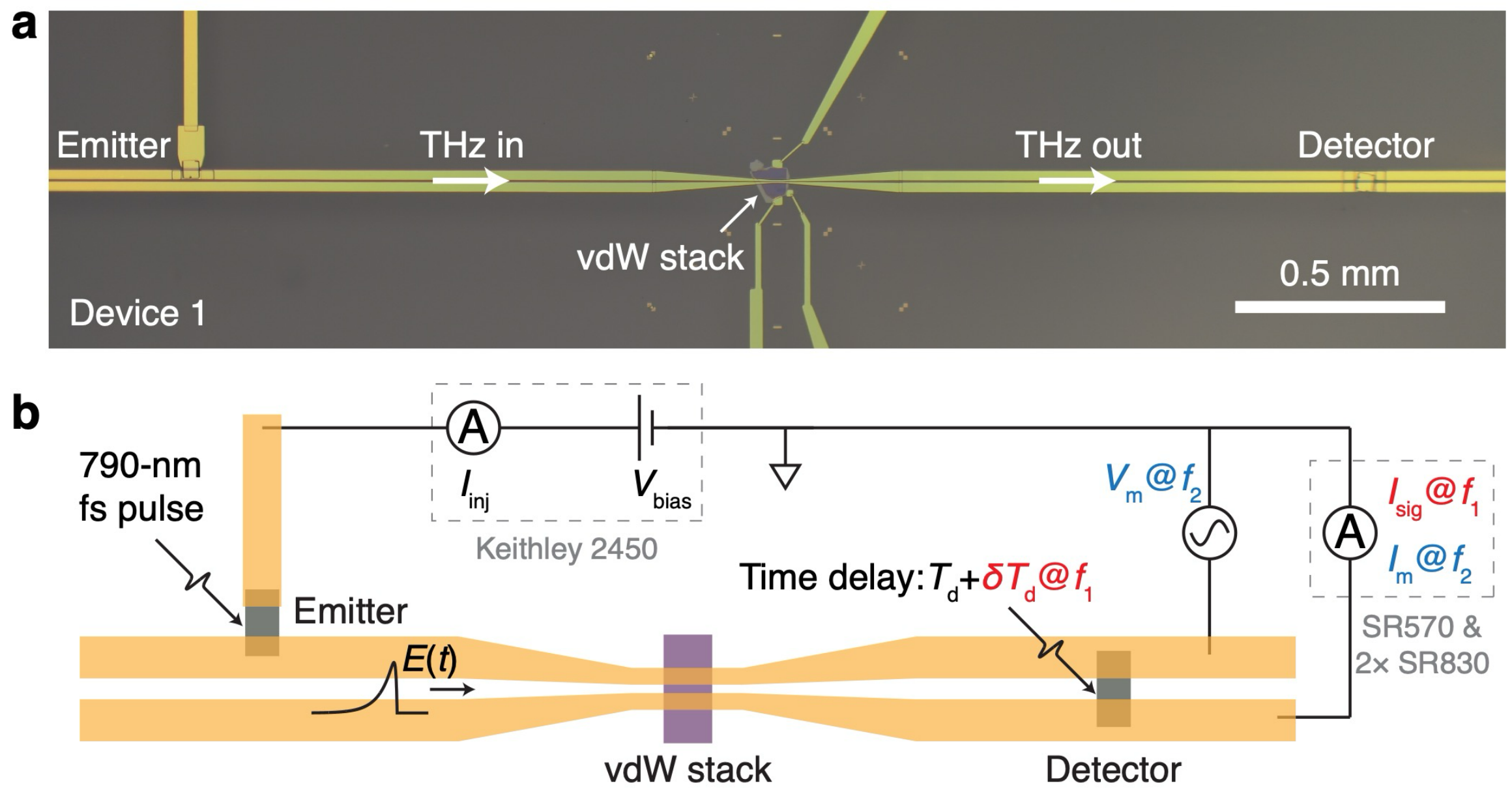

**Extended Data Fig. 1 | Device structure and measurement scheme. a**, Optical micrograph of Device 1. Photoconductive switches (emitter and detector) are placed at trisecting points of the CPS, which extends beyond the field of view. The center section of CPS is tapered with near-constant width-to-gap ratio, which enhances sample absorption, thus measurement sensitivity, without introducing significant multireflection effects. **b**, Schematic of the measurement setup. A femtosecond laser (Coherent Chameleon, 80 MHz, 790 nm) excites first the emitter and after a time delay ($T_d$) the detector. The emitter, DC biased at $V_{bias} \sim 10$ V, generates pulsed THz field ($E$) in the CPS after excitation. The emitter photocurrent $I_{inj}$, proportional to the amplitude of generated THz pulses, is measured using a source meter (Keithley 2450). The detector photocurrent, driven by the transmitted THz field at $T_d$, is collected using a current preamplifier (SR570). We modulate $T_d$ with an amplitude of 0.17 ps at $f_1 = 50$ Hz, and read out the resulting modulation ($I_{sig}$) in detector photocurrent using a lock-in amplifier (SR830), effectively measuring $dE(T_d)/dT_d$. Meanwhile, we also apply a small AC monitoring voltage $V_m = 3$ mV at $f_2 = 127$ Hz across the detector, and use a second lock-in amplifier to record the corresponding photocurrent $I_m$ proportional to the responsivity of the detector. The quantity $I_{sig}/(I_{inj} \cdot I_m)$, which represents the intrinsic transmission largely unaffected by external drifts, is analyzed.

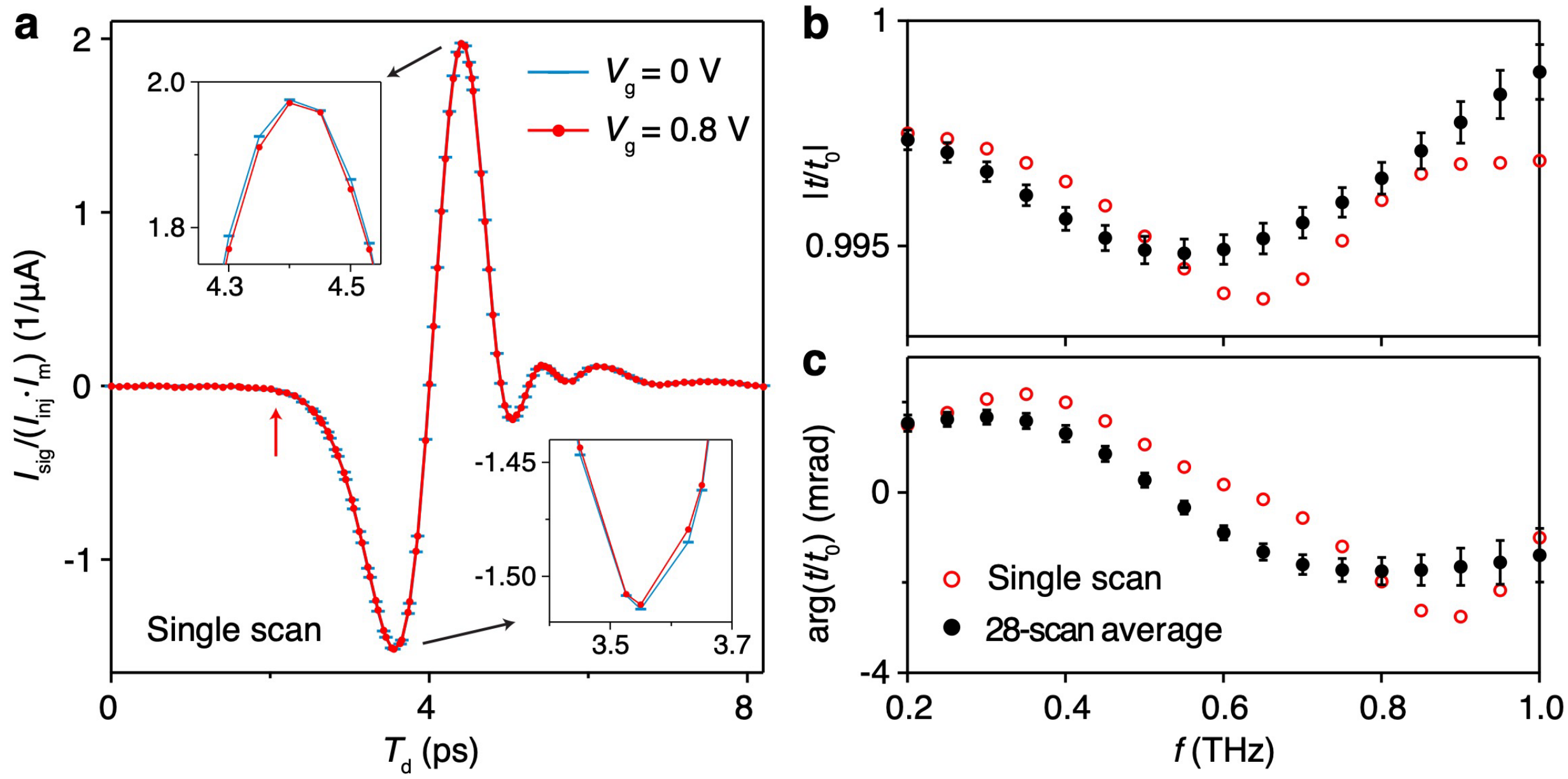

**Extended Data Fig. 2 | Example of time-domain traces and complex transmission. a**, Transmitted signal $I_{sig}/(I_{inj}\cdot I_m)$ in the time domain for undoped (blue, used as reference) and lightly doped (red, same density as Fig. 1f) monolayer $MoSe_2$, measured in a single scan in Device 1. Red arrow approximately marks the onset of signal. Insets show zoomed-in plots of the local extrema of the curves. The gate voltage $V_g$ is set as the fast axis during scans to ensure the $T_d$ values for each pair of curves exactly match. **b**, Red: amplitude of the normalized transmission ($t/t_0$) as a function of frequency obtained using the Fourier transforms of the data in **a**. A Hanning window is applied and the data are zero-padded to a total length of 20 ps before the Fourier transform. Black: amplitude of $t/t_0$ averaged over multiple $t/t_0$ measurements; error bars represent 1σ error from averaging multiple scans. **c**, Same as **b** for the phase of $t/t_0$. The black data points in **b** and **c** are then converted to the data in Fig. 1f using simulated mappings such as those in Extended Data Fig. 3 (see Supplementary Information for data at other gate voltages).

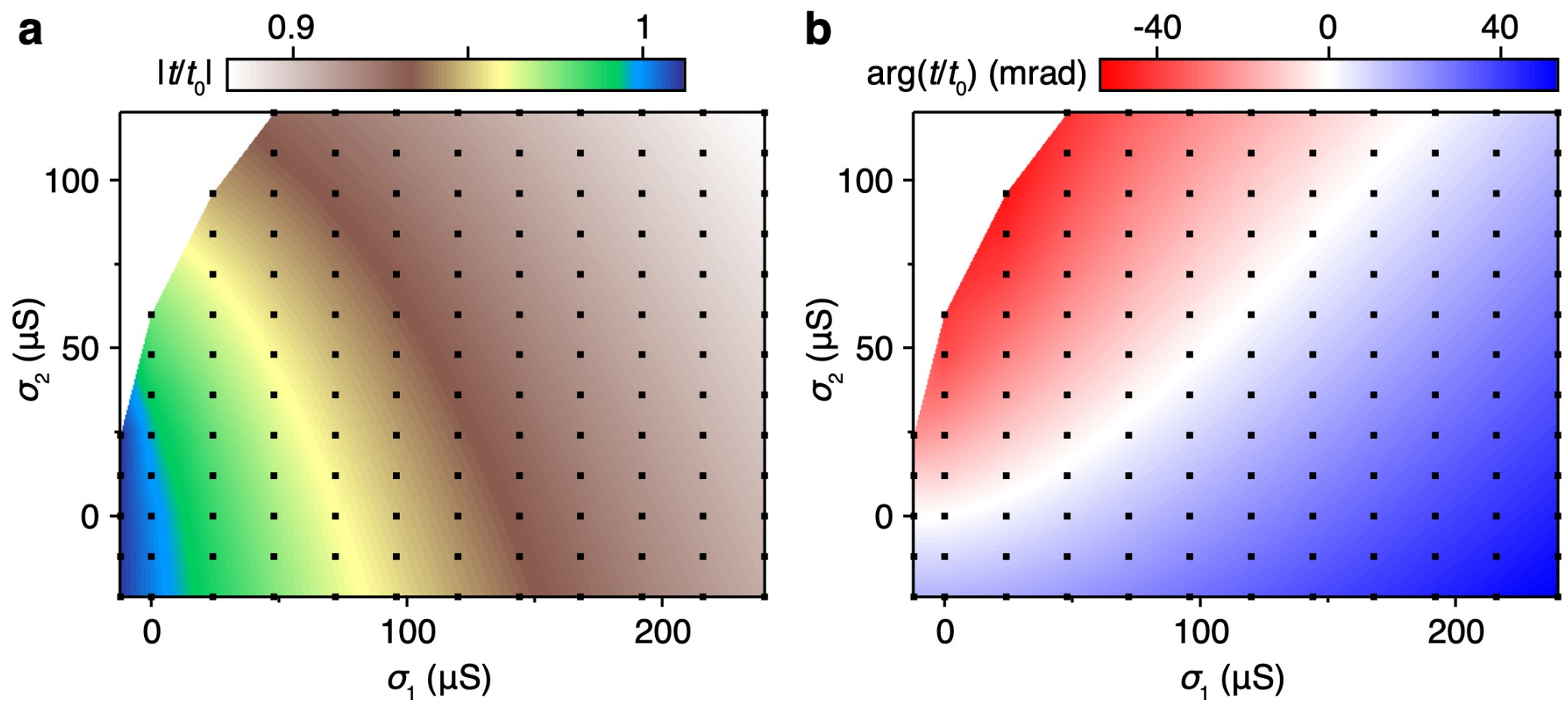

**Extended Data Fig. 3 | Example of mapping between $t/t_0$ and $\sigma$. a,** Amplitude of the transmission coefficient $t$ at 0.2 THz as a function of $\sigma_1$ and $\sigma_2$, normalized by that at $\sigma_1 = \sigma_2 = 0$ ($t_0$). Data are calculated for Device 1. **b,** Same as **a** for the phase shift arg($t/t_0$). To account for the $V_g$ dependence of ITO conductivity ($\sigma_{ITO}$), $t$ is calculated using $\sigma_{ITO}$ = 25.1 μS, the experimentally measured value at the $V_g$ of interest (0.8 V in this case) and 4.5 K, while $t_0$ is calculated using $\sigma_{ITO}$ = 25.9 μS measured at the reference $V_g$ (0 V for Device 1, where $MoSe_2$ is intrinsic). Similar mappings are generated for each frequency, temperature, and $V_g$, to convert the experimentally measured $t/t_0$ to sample $\sigma$. To obtain such mappings, simulations using accurate device geometry (see Ref.[31] for details) are first performed on the $\sigma$ grid marked by black dots with sufficient $\sigma_{ITO}$ steps. The results are then linearly interpolated between $\sigma_{ITO}$ values. A Voronoi interpolation over $\sigma$ is further used to generate the color maps such as those in **a** and **b**.

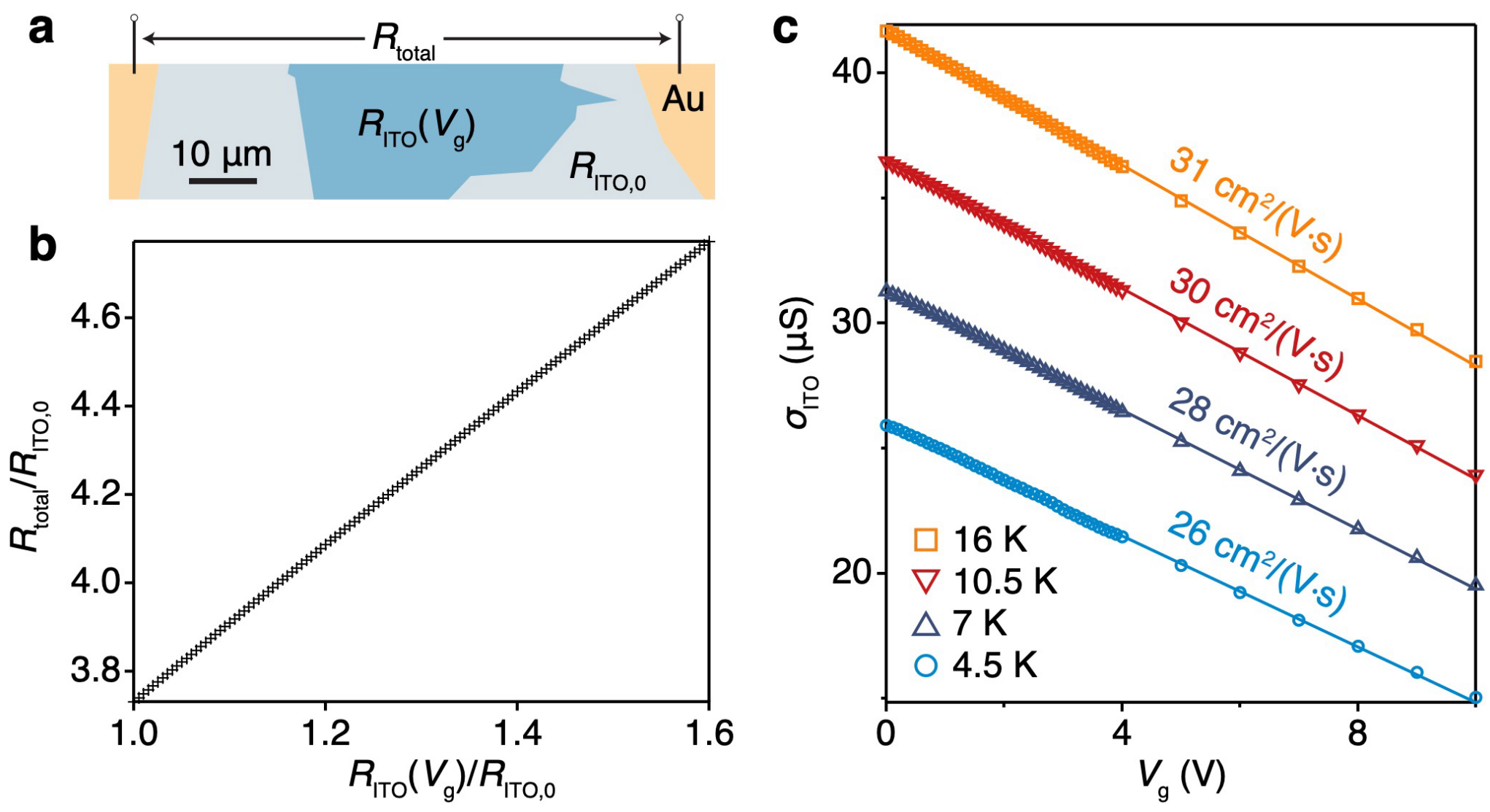

**Extended Data Fig. 4 | ITO conductivity measurement. a**, Accurate geometry of the ITO gate electrode in Device 1 (see also Fig. 1b), obtained using optical microscopy. Darker blue area is electrostatically gated by the $MoSe_2$ or few-layer graphite, and thus exhibits a $V_g$-dependent sheet resistance $R_{ITO}(V_g)$. Lighter blue area has constant sheet resistance $R_{ITO,0}$. The full strip is contacted by Cr/Au on both ends and its total resistance $R_{total}$ is measured as a function of $V_g$, using a Keithley 2450 source meter with a bias voltage smaller than 20 mV. The contact and Au resistances are negligible relative to $R_{ITO}$. **b**, $R_{total}/R_{ITO,0}$ as a function of $R_{ITO}(V_g)/R_{ITO,0}$, calculated using finite element analysis (COMSOL). The calculated curve is used to convert the experimentally measured $R_{total}$ to $R_{ITO}$ and thus $\sigma_{ITO}=1/R_{ITO}$. **c**, $\sigma_{ITO}$ as a function of $V_g$ across different temperatures (markers). Solid lines are linear fits to the data. The electron mobility in ITO is extracted from the linear slope and marked on each line. The mobility values translate to[31,40] a transport scattering time $\tau_{ITO}$ around 5 fs. Thus, $\sigma_{ITO}$ can be treated as frequency independent below a few THz (see also direct measurements from Ref.[41] on ITO thin films with similar mobility values). Leveraging the linear dependence of $\sigma_{ITO}$ on $V_g$ and thus carrier density, the ITO measurement is also used to sense the doping onset in $MoSe_2$ (Extended Data Fig. 6). In addition, the temperature dependence of $\sigma_{ITO}$ allows us to determine the actual sample temperature during the THz measurement (Methods).

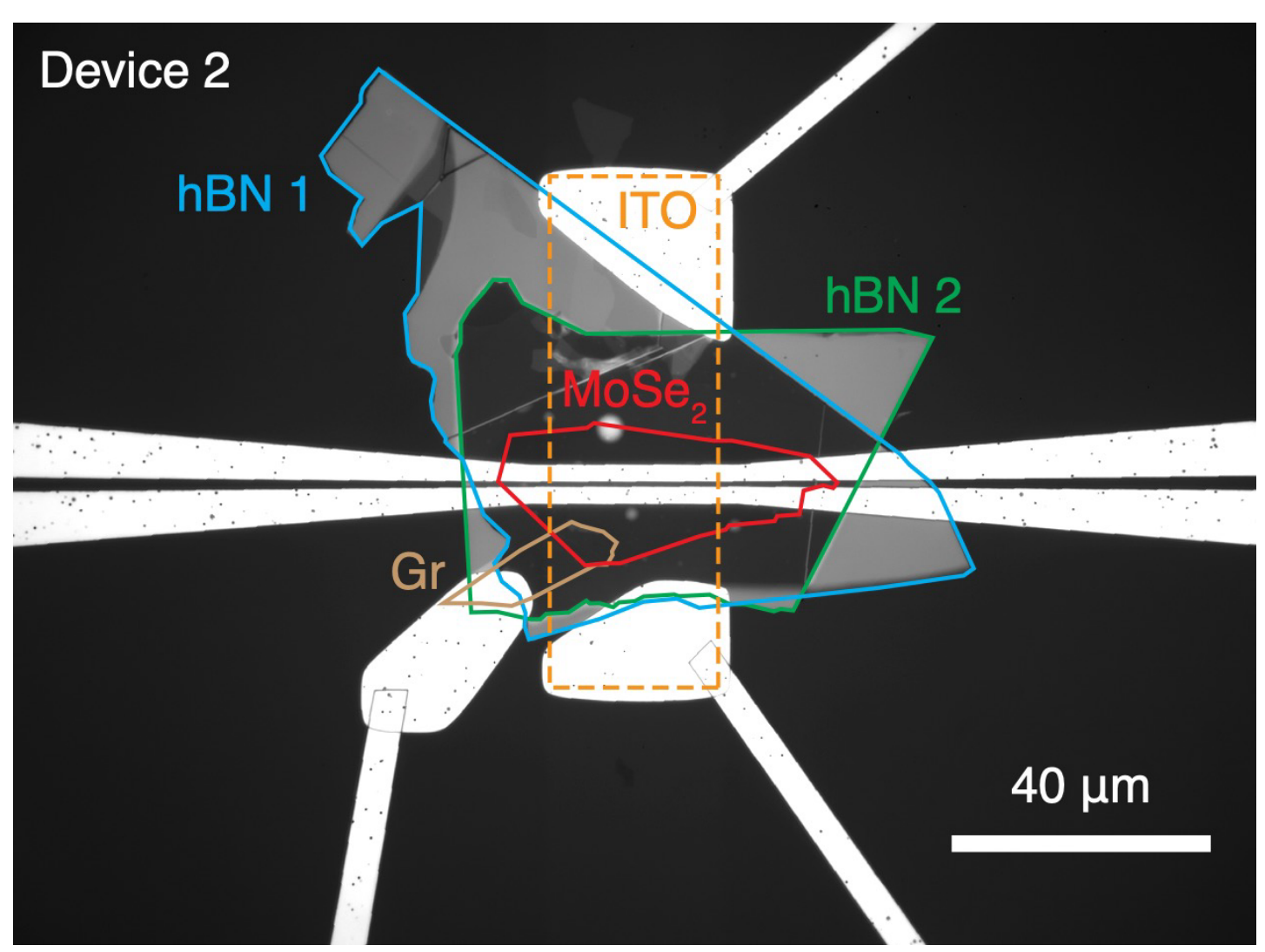

**Extended Data Fig. 5 | Optical micrograph showing center section of the second device.** The flakes are outlined in solid. The yellow dashed box marks the ITO backgate.

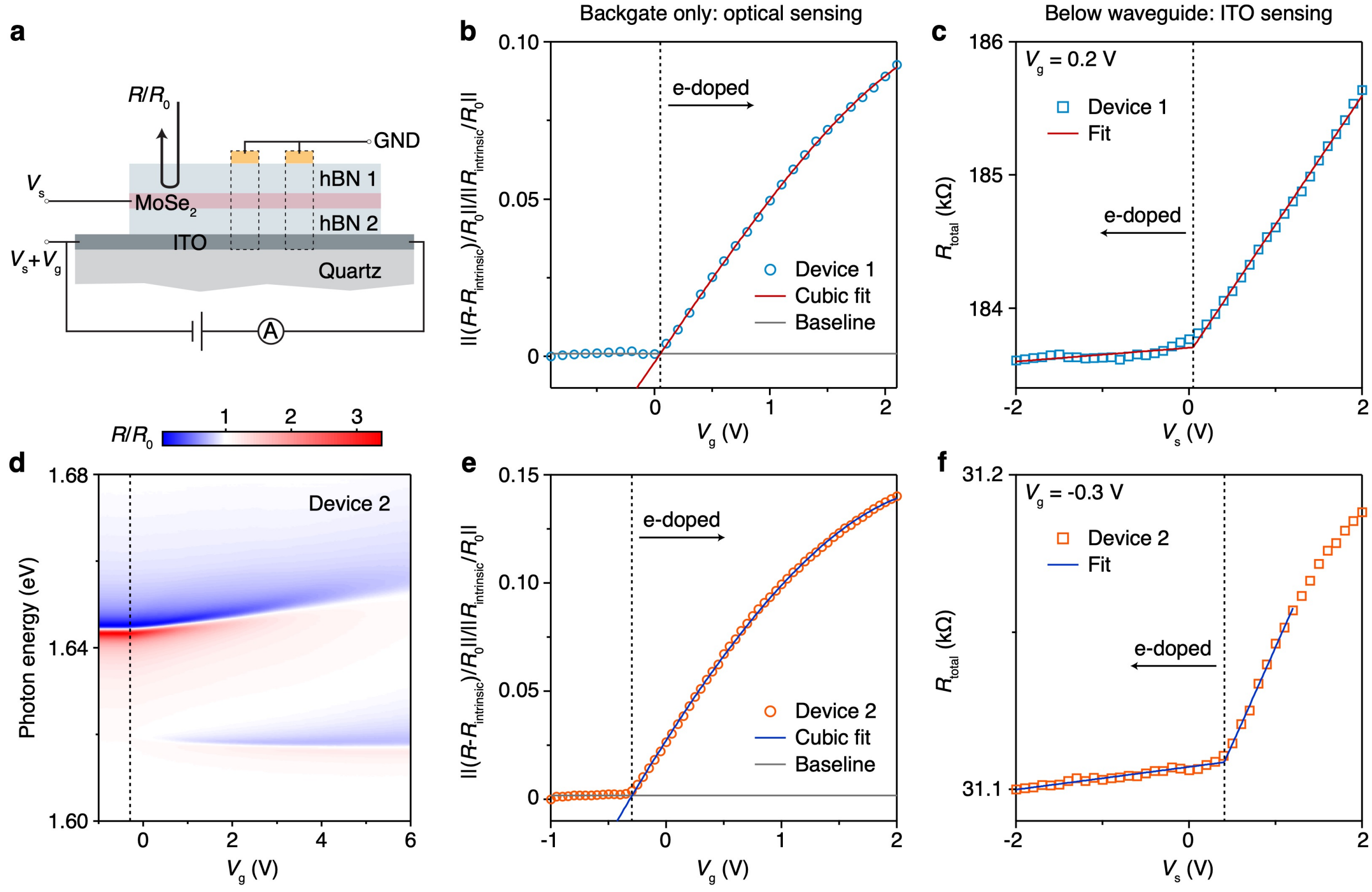

**Extended Data Fig. 6 | Carrier density calibration. a**, Cross-sectional schematic of device and measurement circuit. DC voltages $V_s$ and $V_s$+$V_g$, defined with respect to the waveguide, are applied to the $MoSe_2$ and ITO, respectively, using either Keithley 2450 or 2400 source meters. Dual-gated regions in $MoSe_2$ are highlighted by the dashed rectangles with exaggerated height. **b**, Change of reflectance versus $V_g$ obtained from the spectra in Fig. 1d. The change is quantified as $||(R\text{-}R_{intrinsic})/R_0||/||R_{intrinsic}/R_0||$. Here $R_{intrinsic}$ represents the spectrum measured at the lowest $V_g$ where $MoSe_2$ is intrinsic, and ||…|| denotes the L1 norm summed over photon energy between 1.60 and 1.68 eV. The average baseline in the intrinsic region is plotted in grey. After electron doping, the data deviates from the baseline, and the trend is fitted by a cubic polynomial (red)[38]. The crossing point marked by the vertical dashed line is taken as the doping onset $V_{g,onset}$ = 0.05 ± 0.02 V. The uncertainty mainly represents difference between increasing-$V_g$ and decreasing-$V_g$ scans. **c**, Total resistance of ITO as a function of $V_s$, measured at 1.6 K and $V_g$ = 0.2 V, where the backgate-only regions in $MoSe_2$ are slightly doped. With increasing $V_s$, the slope change indicates that $MoSe_2$ in the dual-gated region becomes intrinsic, such that the CPS traces directly gate the ITO[31]. The turning point ($V_T$, vertical dashed line) is determined by fitting the data to $R_{total} = a\cdot\max(0, V_s\text{-}V_T) + b\cdot V_s + c$, where $a$, $b$, and $c$ are the additional fitting parameters. From

these measurements and the measured hBN thickness, we find that $V_s \approx -0.1$ V aligns the density in the dual-gated region to the rest of $MoSe_2$. Thus, we set $V_s$ to -0.1 V for THz measurements on Device 1. Quantum-capacitance corrections are negligible here because of the relatively thick hBN flakes in our devices. **d**, Optical reflectance spectra from Device 2 at 4.8 K. Vertical dashed line marks $V_{g,onset}$. **e**, Same as **b** for Device 2. $V_{g,onset}$ is found to be $-0.30 \pm 0.02$ V. **f**, Same as **c** for Device 2, measured at 4.8 K and $V_g = -0.3$ V. The $V_s$ for THz measurements is found to be 0.4 V.

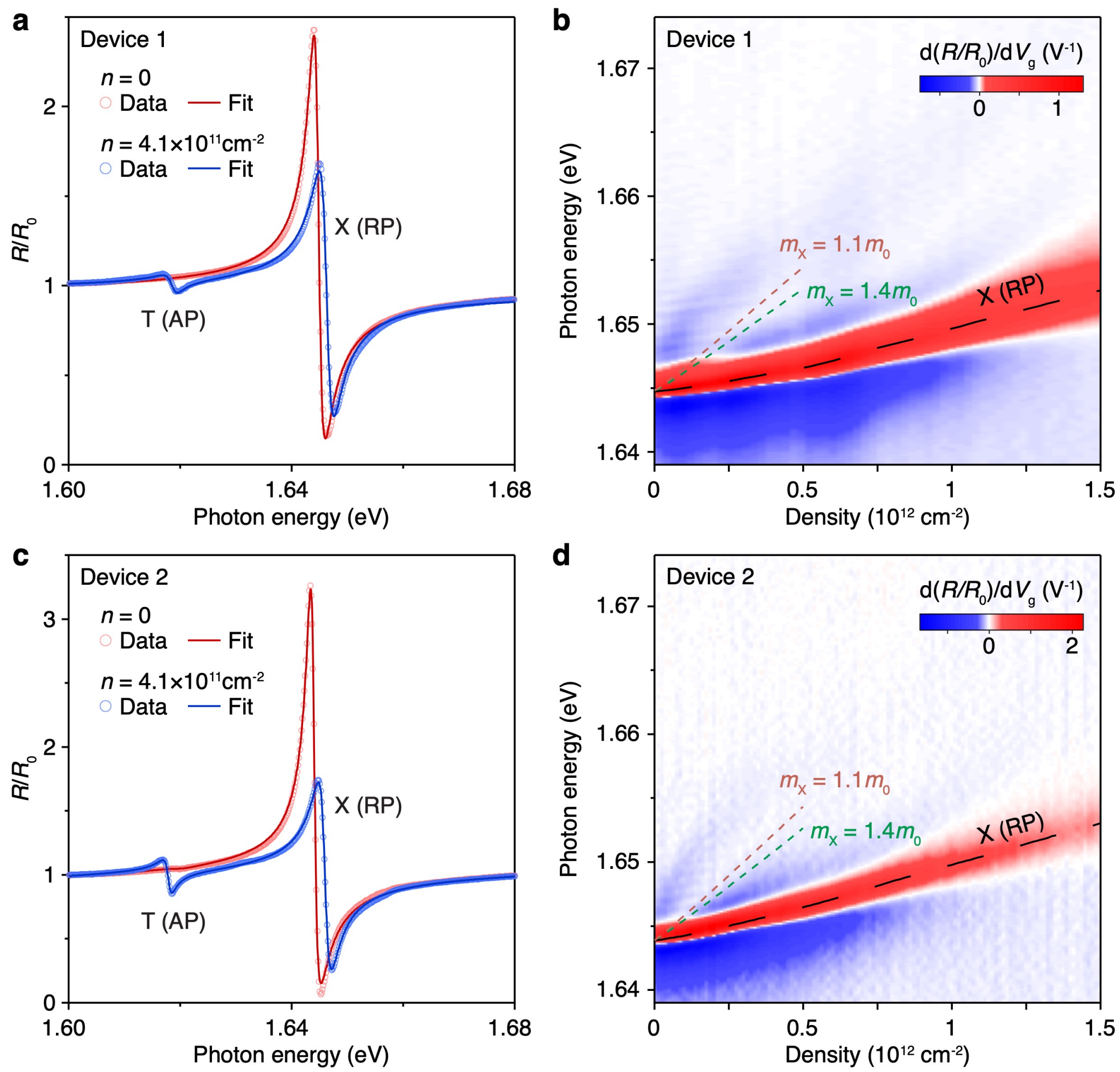

**Extended Data Fig. 7 | Exciton response and Umklapp sideband. a**, Optical reflectance spectra at selected densities from Fig. 1d. To extract the exciton energy, each spectrum is fitted with a Fano resonance plus a linear background (red). For higher densities, a second Fano resonance is included to capture the trion feature (blue). **b**, Same as Fig. 1e, now showing the expected position of Umklapp sideband for two $m_x$ values: $m_x = 1.4\ m_0$ inferred from the previously measured electron effective mass[22] and exciton reduced mass[42]; and $m_x = 1.1\ m_0$ taken from the best fit in Ref.[10]. Both values are consistent with our data. **c**, Same as **a** for Device 2, using data from Extended Data Fig. 6d. The resonances here are slightly sharper than those in Device 1. **d**, Same as **b** for Device 2 showing consistent results. Although the Umklapp feature appears to fade at slightly lower densities, a quantitative comparison between devices is difficult due to its small oscillator strength and the difference in exciton linewidths to begin with.

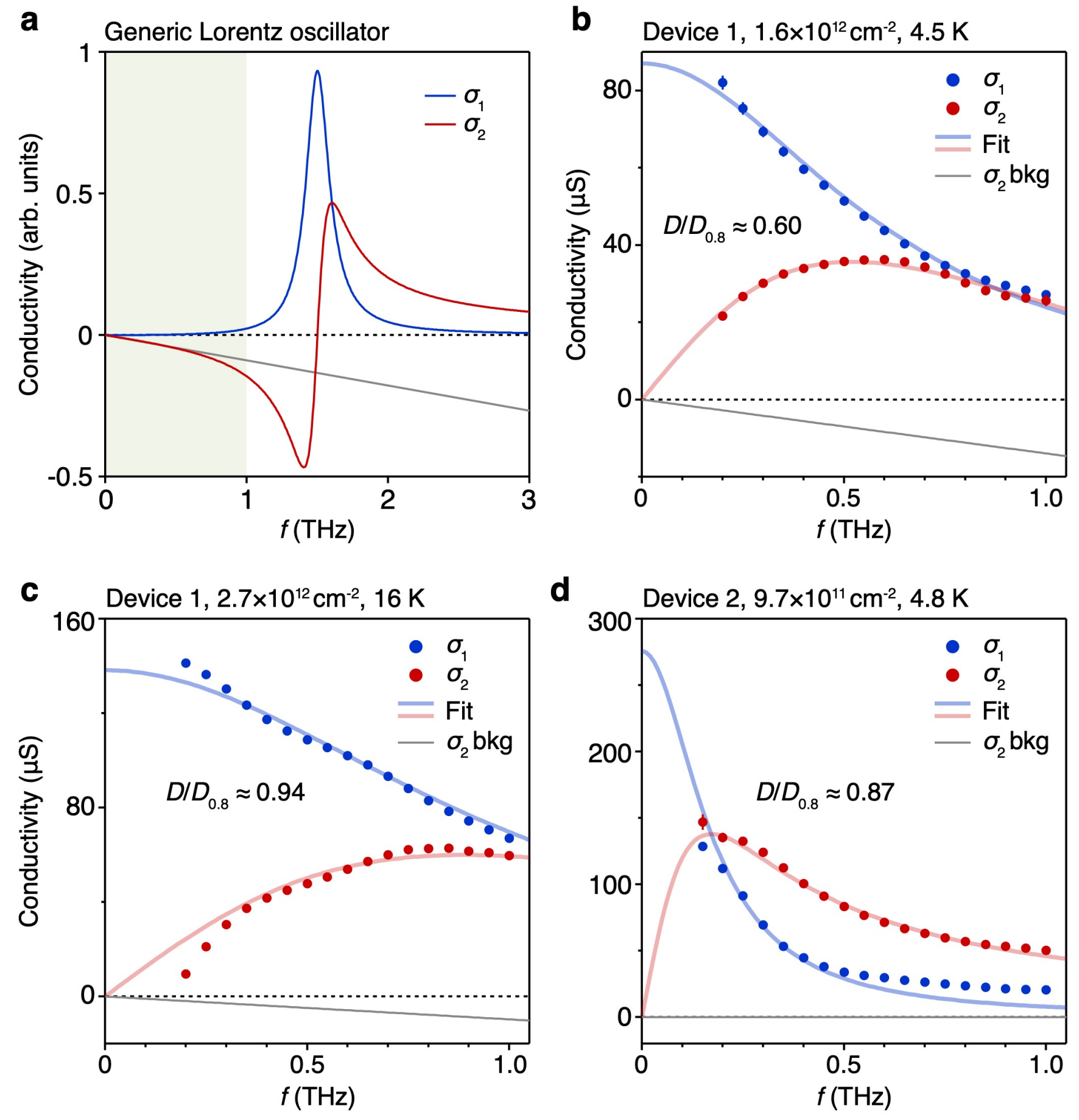

**Extended Data Fig. 8 | Drude weight in the metallic regime. a**, Frequency-dependent conductivity of a generic Lorentz oscillator. At frequencies below the absorption feature (shaded region), although the real part of conductivity is almost zero, the imaginary part remains significant and can be approximated as $-i\alpha f$ (grey line). We note that the same approximation holds for broader absorption features, which can be modeled as linear combinations of Lorentz oscillators. **b**, Same as Fig. 1g with the fitted $-i\alpha f$ term ($\sigma_2$ bkg) also plotted. The ratio between the fitted ($D$) and expected Drude weight with $m^* = 0.8\, m_0$ ($D_{0.8}$) is labeled in the panel. Both the notable deviation of $D/D_{0.8}$ from unity and the significant $-i\alpha f$ term suggest spectral weight transfer from the Drude peak to absorption features above our frequency window. **c**, Same as **b** but at 16 K and $2.7\times10^{12}$ cm$^{-2}$. $D/D_{0.8}$ recovers to near 1, and the $-i\alpha f$ term also weakens. **d**, Same as Fig. 2b inset with the $-i\alpha f$ term plotted. $D/D_{0.8}$ is found to be near unity and $\alpha$ is found to be zero under the non-negativity constraint.

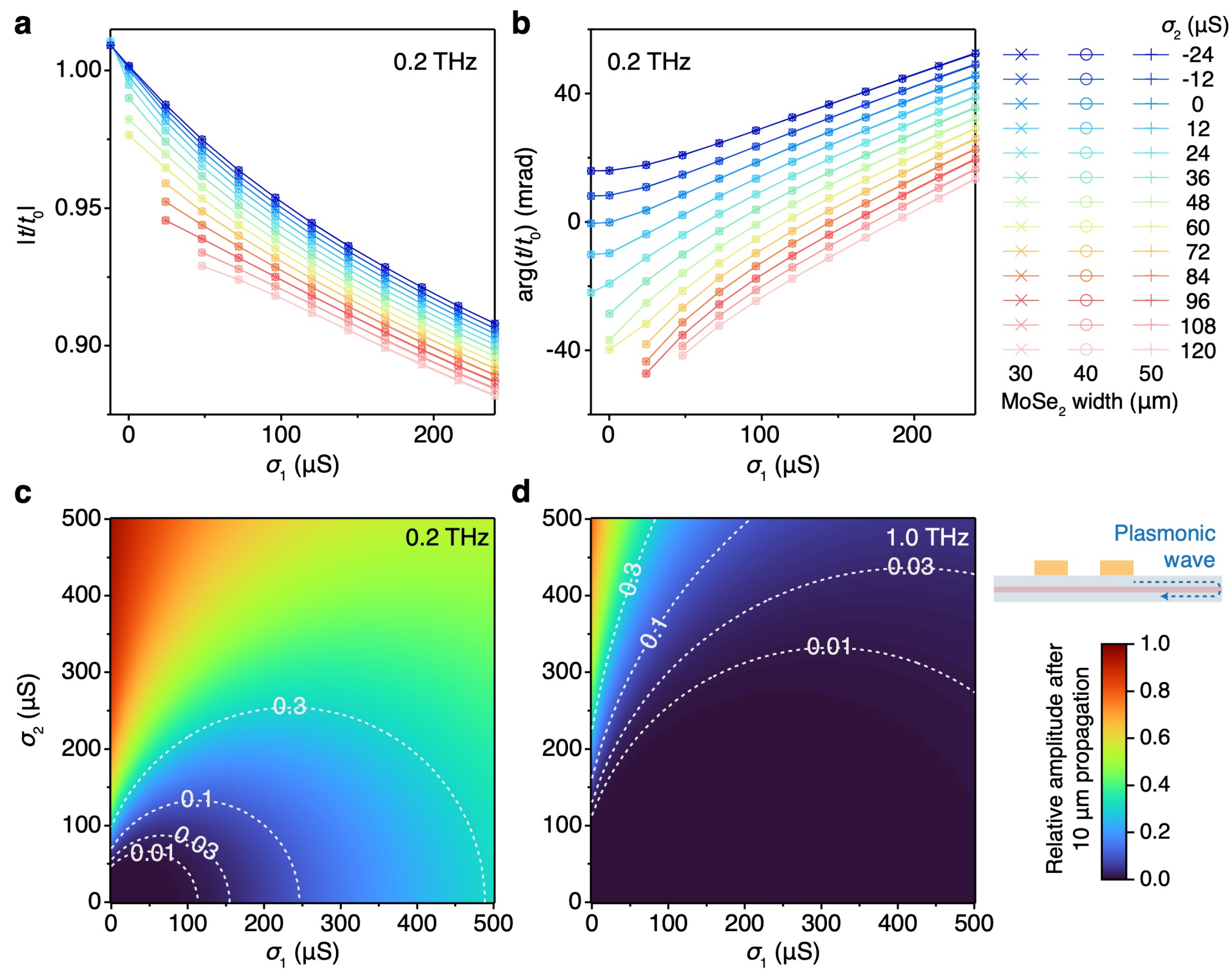

**Extended Data Fig. 9 | Evaluating plasmonic effects associated with finite sample width. a,** Simulated $|t/t_0|$ at 0.2 THz (same as those in Extended Data Fig. 2a) as a function of $\sigma_1$. Each color represents a distinct $\sigma_2$ value, as labeled on the right. Crosses, circles, and plus markers denote total $MoSe_2$ cross-sectional widths of 30, 40, and 50 μm used in three different sets of simulations, respectively. **b**, Same as **a** for $\arg(t/t_0)$. The lack of width dependence indicates that within the conductivity range considered, plasmonic effects associated with finite sample width are negligible. **c**, Estimated relative amplitude of a plasmonic wave at 0.2 THz after propagating 10 μm away from the CPS with $\sigma_{ITO}$ = 20 μS (see Methods for details). The decay exceeds 10-fold in the conductivity range studied, which explains the absence of width dependence in **a** and **b**. **d**, Same as **c** at 1.0 THz, demonstrating that the decay is even more pronounced at higher frequencies for fixed conductivity values. We remark that the plasmonic effects will become relevant at very high densities not studied in this work, where the conductivity values fall into the insufficiently damped regions (green to orange) in **c** and **d**.

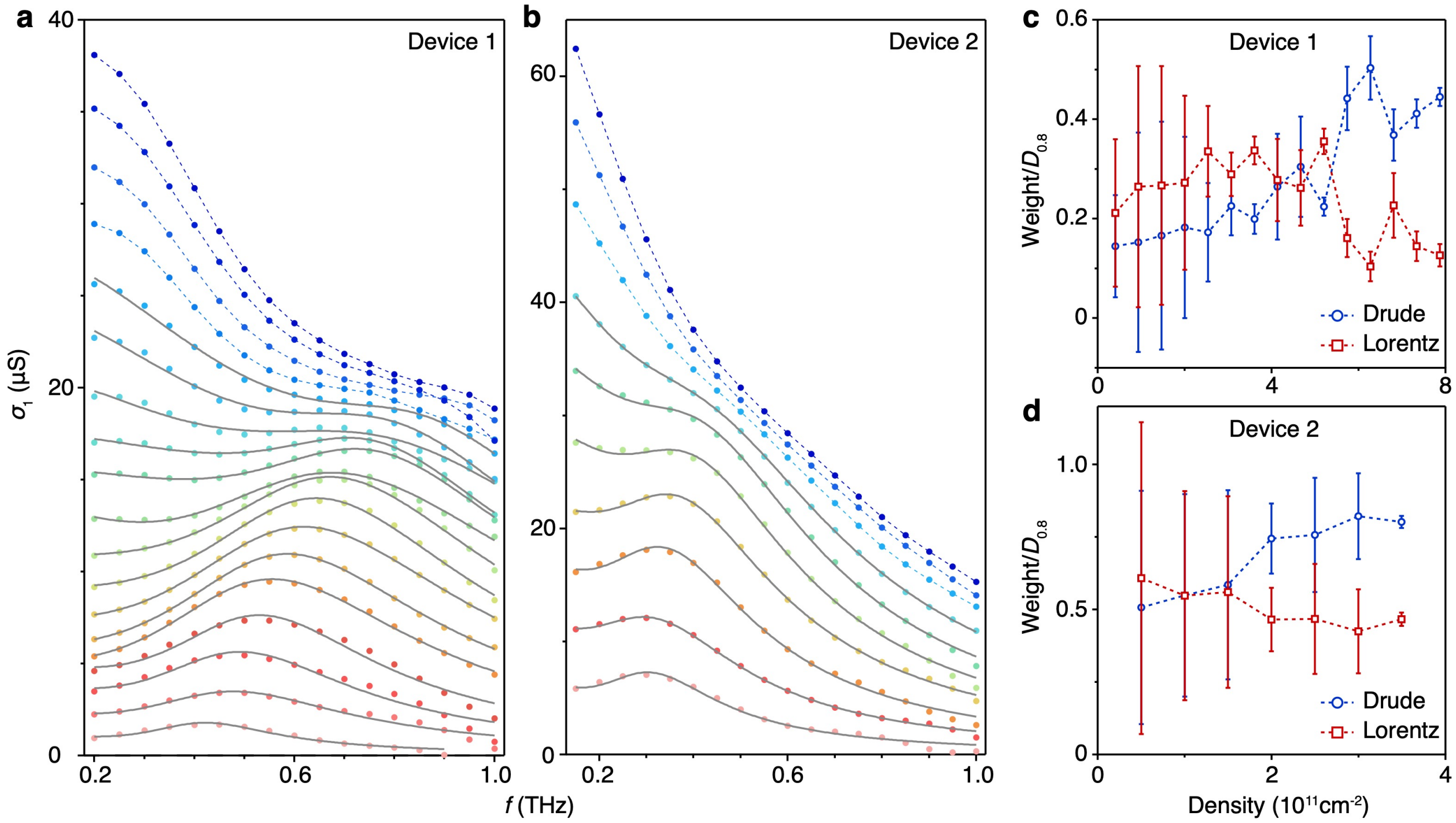

**Extended Data Fig. 10 | Fitting conductivity spectra.** **a**, Same as Fig. 2a, now shown with the fitted curves (grey). The top 4 curves are excluded from the fitting because the pinning mode cannot be reliably identified. **b**, Same as **a** for Device 2, using data from Fig. 2b. Here, the top 3 curves are excluded for the same reason. **c,** Drude and Lorentz weights extracted from **a**, normalized by the expected Drude weights ($D_{0.8}$) of an electron gas with $m^* = 0.8\ m_0$ at corresponding densities. Error bars represent 1σ error from fitting. While the extracted position and width of the Lorentz component are relatively accurate (as indicated by the small error bars in Fig. 3c), the weights shown here have significant uncertainties. Nevertheless, the data reveal the trend that the Drude component overtakes the Lorentz component with increasing density. **d**, same as **c** for Device 2. We remark that fitting both real and imaginary parts of $\sigma$, with an added $-i\alpha f$ background term, yields consistent but slightly more scattered results.

**Supplementary Information for**

**Terahertz electrodynamics in a zero-field Wigner crystal**

Su-Di Chen[1,2,3,4*], Ruishi Qi[1,2,3], Ha-Leem Kim[1,2,3], Qixin Feng[1,2], Ruichen Xia[1], Dishan Abeysinghe[1,2], Jingxu Xie[1,2,5], Takashi Taniguchi[6], Kenji Watanabe[7], Dung-Hai Lee[1,2], Feng Wang[1,2,3*]

[1]Department of Physics, University of California, Berkeley, CA 94720, USA.

[2]Materials Sciences Division, Lawrence Berkeley National Laboratory, Berkeley, CA 94720, USA.

[3]Kavli Energy NanoScience Institute, University of California, Berkeley and Lawrence Berkeley National Laboratory, Berkeley, CA 94720, USA.

[4]Department of Physics, University of Florida, Gainesville, FL 32611, USA.

[5]Graduate Group in Applied Science and Technology, University of California, Berkeley, CA 94720, USA.

[6]Research Center for Materials Nanoarchitectonics, National Institute for Materials Science, 1-1 Namiki, Tsukuba 305-0044, Japan.

[7]Research Center for Electronic and Optical Materials, National Institute for Materials Science, 1-1 Namiki, Tsukuba 305-0044, Japan.

[*]Corresponding authors. Email: sudichen@ufl.edu, fengwang76@berkeley.edu

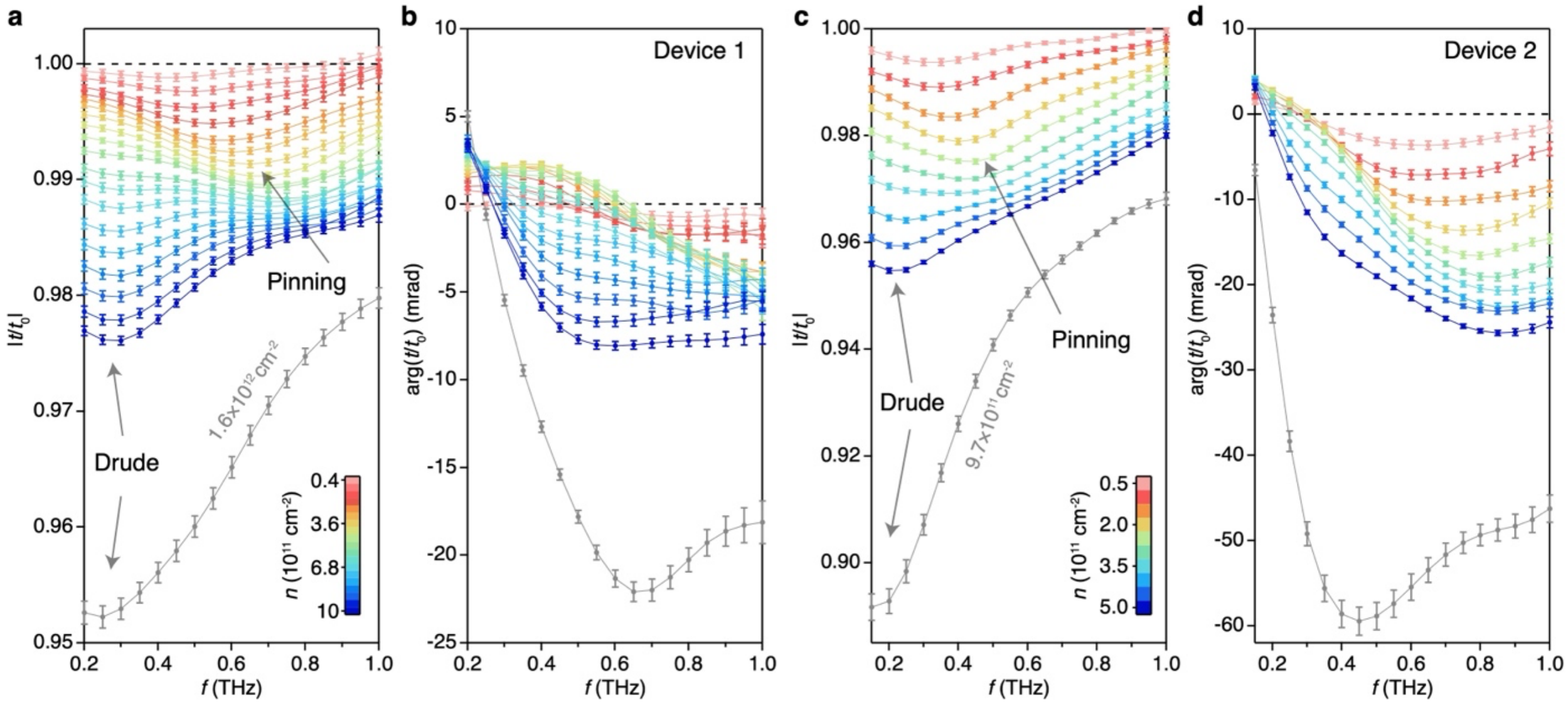

**Fig. S1. Density evolution of transmission spectra ($t$) normalized by the reference transmission ($t_0$).** **a**. Amplitude of $t/t_0$ as a function of frequency at various gate voltages from Device 1. The colored curves underlie the data in Fig. 2a; the grey curve underlies the data in Fig. 1g. Arrows mark the features that give rise to the Drude peak and pinning mode in the $\sigma$ spectra after numerical mapping. **b**. Same as a for the phase of $t/t_0$. **c, d**, same as a, b for Device 2. The colored curves underlie the data in Fig. 2b, and the grey curves underlie the data in its inset.

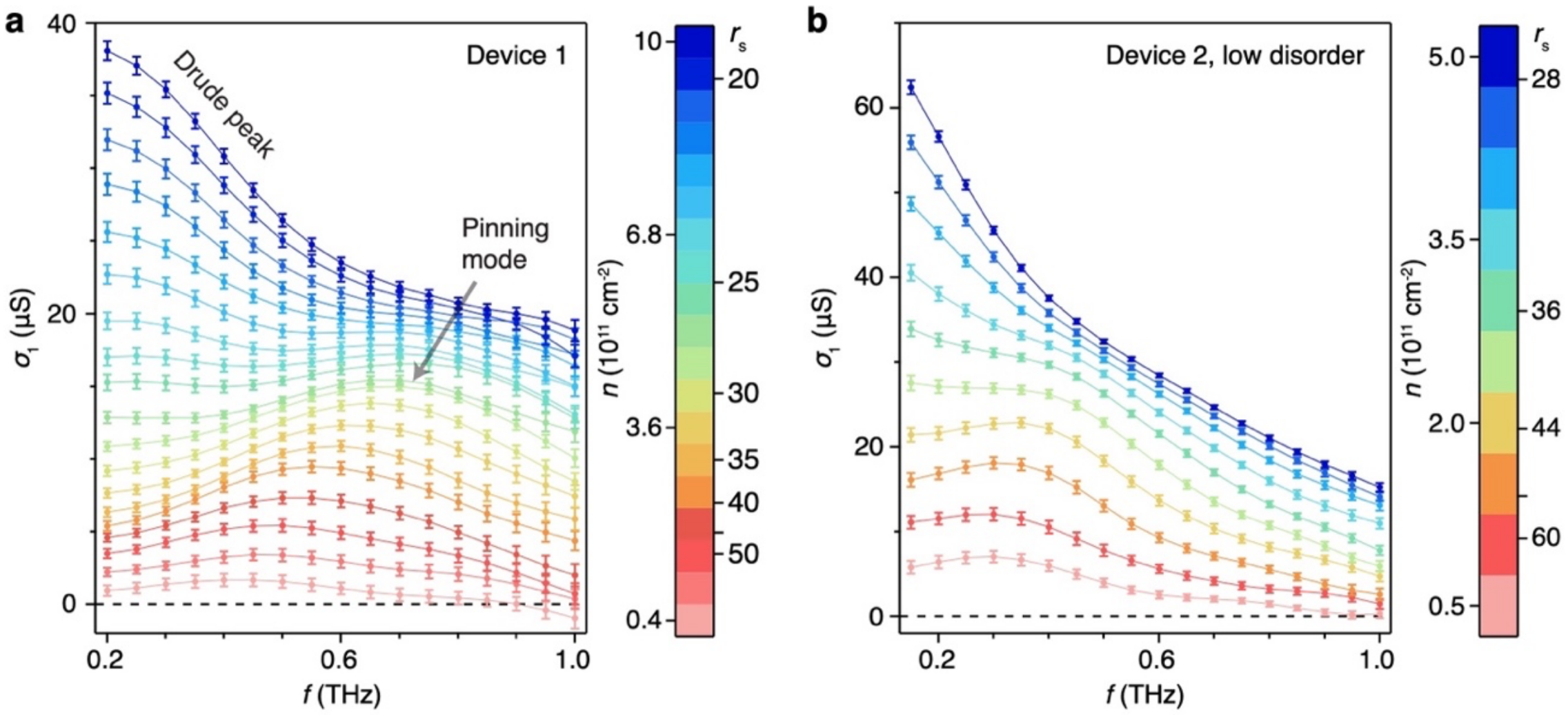

**Fig. S2. Density and disorder dependence of AC conductivity.** Data are the same as those plotted in Fig. 2a and b, but with error bars included. The error bars represent 1σ uncertainty from averaging multiple measurements. For Device 1, the data points at 0.95 and 1 THz at the lowest density (bottom curve in a) are excluded from the analysis in the main text because the signal-to-noise ratio therein is insufficient to yield physical values of $\sigma_1$.